# A Transform Approach to Linear Network Coding for Acyclic Networks with Delay

Teja Damodaram Bavirisetti, G. Abhinav, K. Prasad and B. Sundar Rajan
Dept. of ECE, IISc, Bangalore 560012, India, Email: {tdamodar, abhig_88, prasadk5, bsrajan}@ece.iisc.ernet.in

*Abstract*—The algebraic formulation for linear network coding in acyclic networks with the links having integer delay is well known. Based on this formulation, for a given set of connections over an arbitrary acyclic network with integer delay assumed for the links, the output symbols at the sink nodes, at any given time instant, is a $\mathbb{F}_q$-linear combination of the input symbols across different generations, where $\mathbb{F}_q$ denotes the field over which the network operates. We use finite-field discrete fourier transform (DFT) to convert the output symbols at the sink nodes, at any given time instant, into a $\mathbb{F}_q$-linear combination of the input symbols generated during the same generation. We call this as transforming the acyclic network with delay into $n$-*instantaneous networks* ($n$ is sufficiently large). We show that under certain conditions, there exists a network code satisfying sink demands in the usual (non-transform) approach if and only if there exists a network code satisfying sink demands in the transform approach. Furthermore, we show that the transform method (along with the use of alignment strategies) can be employed to achieve half the rate corresponding to the individual source-destination min-cut (which are assumed to be equal to 1) for some classes of three-source three-destination unicast network with delays, when the zero-interference conditions are not satisfied.

## I. INTRODUCTION

Network coding was introduced in [1] as a means to improve the rate of transmission in networks. Linear network coding was introduced in [2] and it was found to be sufficient to achieve the maxflow-mincut capacity in certain scenarios such as multicast. The existence problem of network coding for networks without delay was converted into an algebraic problem in [3]. The case of acyclic networks with delays was abstracted in [3] as acyclic networks where each link in the network has an integer delay associated with it.

The problem of network coding for multiple unicast sessions was considered in [4], [5]. In [6], the concept of *interference alignment* from interference channels [7] was extended to instantaneous unicast networks with three source-destination pairs for the case where, each source-destination pair has a min-cut of 1. This was called *network alignment* and it is useful in guaranteeing a minimum throughput when the zero-interference conditions in Theorem 6 of [3] cannot be satisfied.

The motivation behind this work is striving to provide a minimum throughput guarantee when the zero-interference conditions cannot be satisfied in an acyclic network with delay, while not making use of any memory at the intermediate nodes (i.e., nodes other than the sources and sinks). The set of all $\mathbb{F}_q$-symbols generated by the sources at any particular time instant are said to constitute the same *generation*. The output symbols at the sink nodes, at any given time instant, is a $\mathbb{F}_q$-linear combination of the input symbols across different generations, where $\mathbb{F}_q$ denotes the field over which the network operates. We convert the output symbols at the sink nodes, at any given time instant, into a $\mathbb{F}_q$-linear combination of the input symbols generated during the same generation, by using techniques similar to Multiple Input Multiple Output-Orthogonal Frequency Division Multiplexing (MIMO-OFDM) [8]. We call this technique as the *transform technique*, since we use DFTs over finite fields towards achieving this instantaneous behaviour in the network. As a first step towards guaranteeing a minimum throughput when the zero-interference conditions cannot be satisfied in an acyclic network with delay, we consider a three-source three-destination unicast network with the source-destination pair denoted as $S_i$-$D_i$ ($i \in \{1,2,3\}$). We also assume a min-cut of one between source $S_i$ and destination $D_i$. Under this set-up, we apply the transform techniques and network-alignment to find conditions under which the network can guarantee a throughput close to half for every source-destination pair $S_i$-$D_i$ ($i \in \{1,2,3\}$). This method does not make use of memory at the intermediate nodes.

The contributions of this paper are as follows.

- We convert the output symbols at the sink nodes, at any given time instant, into a $\mathbb{F}_q$-linear combination of the input symbols generated during the same generation using finite-field Discrete Fourier Transform (DFT). We call this as transforming the acyclic network with delay into $n$-*instantaneous networks*, where, $n$ is sufficiently large.
- Using a constructive proof, we show that there exists a network code (satisfying a certain property) that achieves the sink demands in the usual (non-transform) approach if and only if there exists a network code satisfying sink demands in the transform approach .
- For a three source-three destination unicast network with delays, which do not satisfy the zero-interference conditions, we extend the transform techniques to achieve atleast half the rate corresponding to the individual source-destination min-cut (which are assumed to be equal to 1), along with the use of alignment strategies. In particular, the contributions for the three source-three destination unicast network with delays are as follows.
  1) When the min-cut between $S_i$-$D_j$ is greater than or equal one, $\forall$ $(i,j) \in \{1,2,3\}$ ($i \neq j$), we derive sufficient conditions under which network alignment can achieve half the rate corresponding to the individual source-destination min-cut, with time-invariant Local Encoding Kernels (LEKs).
2) The network alignment procedure with time-invariant LEKs is then generalized with the use of time-varying LEKs.
3) When the min-cut between $S_i$-$D_j$ is zero for some $(i,j) \in \{1,2,3\}$ ($i \neq j$), we derive sufficient conditions under which network alignment can achieve atleast half the rate corresponding to the individual source-destination min-cut.

The organization of this paper is as follows. In Section II, we review the system model for acyclic networks with delays presented in [3]. Section III presents the central contribution of this work, i.e., the transform technique using which we convert the usual convolutional behaviour of the network into instantaneous behaviour. In Section III, we also prove the interchangeability of solving the usual (non-transform) network code existence problem and the counterpart in the transform technique. In Section IV, we combine our transform technique with the alignment techniques for acyclic instantaneous networks given in [6] to achieve an asymptotic throughput of $1/2$ for certain classes of acyclic networks with delays, even when the zero-interference conditions cannot be satisfied in such networks. We conclude our paper in Section V with a discussion and directions for further research.

*Notations:* The cardinality of a set $E$ is denoted by $|E|$. A superscript of $t$ accompanying any variable (for example, $\epsilon^{(t)}$) or any matrix (for example, $M^{(t)}$) denotes that they are a function of time $t$. The $i^{\text{th}}$ row, $j^{\text{th}}$ column element of a matrix $A$ is denoted by $[A]_{ij}$. The notation $P \subset Q$ denotes that the columns of the matrix $P$ are a subset of the columns of the matrix $Q$. Span$(P)$ indicates the sub-space spanned by the columns of the matrix $P$. The determinant of a square matrix $A$ is denoted by $det(A)$. An identity matrix of size $\mu \times \mu$ is denoted by $I_\mu$. For three-source three-destination unicast networks we shall use the term destination to denote sink. A Galois Field of cardinality $p^m$ is denoted by $GF(p^m)$ where, $p$ is a prime number and $m$ is a positive integer.

## II. SYSTEM MODEL

First, we shall briefly review the system model from [3]. We consider a network represented by a Directed Acyclic Graph (DAG) $\mathcal{G} = (V, E)$, where $V$ is the set of nodes and $E$ is the set of directed links. We assume that every directed link between a pair of nodes represents an error-free link and has a capacity of one $\mathbb{F}_q$ symbol per link-use. Multiple links between two nodes are allowed and the $i^{th}$ directed link from $v_1 \in V$ to $v_2 \in V$ is denoted by $(v_1, v_2, i)$. The head and tail of a link $e = (v_1, v_2, i)$ are denoted by $v_2 = \text{head}(e)$ and $v_1 = \text{tail}(e)$. A link between a pair of nodes can have an arbitrary finite integer delay. Let $\mathcal{X}(v) = \{X(v,1), X(v,2), ..., X(v,\mu_v)\}$ be the collection of discrete random processes that are generated at the node $v$. Let $\underline{X}_v = [X(v,1)\ X(v,2)\ ...\ X(v,\mu_v)]^T$. The random process transmitted through link $e$ is denoted by $Z(e)$. Communication is to be established between selected nodes in the network, i.e., we are required to replicate a subset of the random process in $\mathcal{X}(v)$ at some different node $v'$. A connection $c$ is defined as a triple $(v, v', \mathcal{X}(v,v')) \in V \times V \times \mathcal{P}_{\mathcal{X}(v)}$, where $\mathcal{P}_{\mathcal{X}(v)}$ denotes the power-set of $\mathcal{X}(v)$. For the connection $c$, $v$ is called the source and $v'$ is called the sink of $c$, i.e., $v = \text{source}(c)$ and $v' = \text{sink}(c)$ (source$(c) \neq \text{sink}(c)$). The collection of $\nu_{v'}$ random processes $\mathcal{Y}(v') = \{Y(v',1), Y(v',2), ..., Y(v',\nu_{v'})\}$ denotes the output at sink $v'$. Let $\underline{Y}_{v'} = [Y(v',1)\ Y(v',2)\ ...\ Y(v',\nu_{v'})]^T$.

The input random processes $X(v,i)$, output random processes $Y(u,j)$ and random processes $Z(e)$ transmitted on the link $e$ are considered as a power series in a delay parameter $D$, i.e., $X(v,i) = \sum_{t=0}^{\infty} X^{(t)}(v,i)D^t$, $Y(u,j) = \sum_{t=0}^{\infty} Y^{(t)}(u,j)D^t$, and $Z(e) = \sum_{t=0}^{\infty} Z^{(t)}(e)D^t$.

Let $\mathcal{G} = (V, E)$ be an acyclic network with arbitrary finite integer delay on its links. $\mathcal{G}$ is a $\mathbb{F}_q$-linear network [3], if for all links the random process $Z(e)$ on a link $e = (v, u, i) \in E$ satisfies

$$Z^{(t+1)}(e) = \sum_{j=1}^{\mu_v} \alpha_{j,e} X^{(t)}(v,j) + \sum_{e':\text{head}(e')=\text{tail}(e)} \beta_{e',e} Z^{(t)}(e')$$

where, $\alpha_{j,e}$ and $\beta_{e',e}$ belong to $\mathbb{F}_q$, where $q = p^m$, for some prime number $p$ and positive integer $m > 0$. The output at any sink node $v'$, is taken to be

$$Y^{(t+1)}(v',j) = \sum_{e':\text{head}(e')=v'} \epsilon_{e',j} Z^{(t)}(e') \quad (1)$$

where $\epsilon_{e',j} \in \mathbb{F}_q$. The coefficients, $\alpha_{j,e}, \beta_{e',e}$ and $\epsilon_{e',j}$ are also called *local encoding kernels* (LEKs). The vector consisting of all LEKs is denoted by $\underline{\varepsilon}$. Note that in [3], the definition for the output processes at any given time instant at any sink involves linear combinations of the received processes and output processes across different previous time instants, and hence the variables involved in such linear combinations together performed the function of decoding the received processes at the sinks to the demanded input processes. However, in (1), at every sink, we only define a preprocessing of the received symbols corresponding to the previous time instant alone. The outputs $Y^{(t+1)}(v',j)$ as $t$ varies, will then be used by sink-$j$ to decode the demanded input processes using sufficient delay elements for feed-forward and feedback operations. These LEKs are time-invariant unless mentioned otherwise.

We assume some ordering among the sources so that the random process generated by the sources can be denoted, without loss of generality, as $\underline{X}_1(D), \underline{X}_2(D), ..., \underline{X}_s(D)$, where $s$ denotes the number of sources and $\underline{X}_i(D)$ is a $\mu_i \times 1$ column vector given by

$$\underline{X}_i(D) = [X_{i1}(D)\ X_{i2}(D)\ ...X_{i\mu_i}(D)]^T.$$

Similarly, we assume some ordering among the sinks so that the output random process at the sinks can be denoted, without loss of generality, as $\underline{Y}_1(D), \underline{Y}_2(D), ..., \underline{Y}_r(D)$, where $r$ denotes the number of sinks and $\underline{Y}_i(D)$ is a $\nu_i \times 1$ column vector given by

$$\underline{Y}_i(D) = [Y_{i1}(D)\ Y_{i2}(D)\ ...Y_{i\nu_i}(D)]^T.$$

Let
$$Y(D) = [\underline{Y_1}(D)^T \ \underline{Y_2}(D)^T \ ... \ \underline{Y_r}(D)^T]^T$$
$$= [y_1(D) \ y_2(D) \ ... \ y_\nu(D)]^T.$$

Also, let
$$X(D) = [\underline{X_1}(D)^T \ \underline{X_2}(D)^T \ ... \ \underline{X_s}(D)^T]^T$$
$$= [x_1(D) \ x_2(D) \ ... \ x_\mu(D)]^T,$$

where $\mu = \sum_{i=1}^{s} \mu_i$ and $\nu = \sum_{i=1}^{r} \nu_i$. Henceforth, the tail of an edge originating from a source will be identified by the source number and the head of an edge terminating at a sink will be identified by the sink number. From [3], we have

$$Y(D) = M(D)X(D) \quad (2)$$

where, $M(D)$ denotes the *network transfer matrix* of size $\nu \times \mu$ with elements from $\mathbb{F}_q[D]$, the ring of polynomials in variable $D$ with coefficients from $\mathbb{F}_q$. Now, $M(D)$ can also be written as

$$M(D) = \begin{bmatrix} M_{11}(D) & M_{21}(D) & \cdots & M_{s1}(D) \\ M_{12}(D) & M_{22}(D) & \cdots & M_{s2}(D) \\ \vdots & \vdots & \vdots & \vdots \\ M_{1r}(D) & M_{2r}(D) & \cdots & M_{sr}(D) \end{bmatrix}. \quad (3)$$

where $M_{ij}(D)$ denote the network transfer matrix from source $i$ to sink $j$ and is of size $\nu_j \times \mu_i$. Let $d'_{max}$ and $d'_{min}$ denote the maximum and the minimum of all path delays from source-$i$ to sink-$j$, $\forall$ $(i,j)$, between which a path exists. Let

$$d_{max} = d'_{max} - d'_{min}$$

Hence, $M(D)$ can be written as

$$M(D) = \sum_{d=d'_{min}}^{d'_{max}} M^{(d)} D^d = \left( \sum_{d=0}^{d_{max}} M^{(d)} D^d \right) D^{d'_{min}},$$

where $M^{(d)} \in \mathbb{F}_q^{\nu \times \mu}$ represents the matrix-coefficients of $D^d$ of the polynomial elements of $M(D)$.

Since $D^{d'_{min}}$ just adds a constant additional delay to all the outputs, we can take, without loss of generality, $M(D)$ as

$$M(D) = \sum_{d=0}^{d_{max}} M^{(d)} D^d. \quad (4)$$

Hence, $M_{ij}(D)$ can be alternatively written as

$$M_{ij}(D) = \sum_{d=0}^{d_{max}} M_{ij}^{(d)} D^d. \quad (5)$$

For each sink-$j$, we also define $M_j(D)$ to be the $\nu_j \times \mu$ submatrix of $M(D)$ that captures the transfer function between all the sources and the sink-$j$, i.e.,

$$M_j(D) = [M_{1j}(D) \ M_{2j}(D) \ ... \ M_{sj}(D)]. \quad (6)$$

In the network $\mathcal{G}$, let $\mathcal{C}_j$ denote the set of all connections to sink-$j$. Let $\mathcal{C} = \cup_{j=1}^{r} \mathcal{C}_j$. The following lemma from [3] states the conditions for solvability of acylic networks with delay.

*Lemma 1 ( [3]):* An acyclic network with delay is solvable iff there exists an assignment to the LEKs $\underline{\varepsilon}$ such that the following conditions are satisfied.
1) *Zero-Interference:* $M_{ij}^{(d)}(l_i) = 0$, for all pairs (source-$i$, sink-$j$) of nodes such that (source-$i$, sink-$j$, $\underline{X_i}^{(l_i)}(D)$) $\notin \mathcal{C}_j$ for all $0 \leq d \leq d_{max}$, where $M_{ij}^{(d)}(l_i)$ denotes the $l_i^{\text{th}}$ column of $M_{ij}^{(d)}$ and $\underline{X_i}^{(l_i)}(D)$ denotes the $l_i^{\text{th}}$ element of $\underline{X_i}(D)$.
2) *Invertibility:* For every sink-$j$, the square submatrix $M'_j(D)$ of $M_j(D)$ formed by juxtaposition of the columns of $M_{ij}(D)$ ($\forall$ $i$; $1 \leq i \leq s$) other than those involved in the zero-interference conditions is invertible over $\mathbb{F}_q(D)$, the field of rationals over $\mathbb{F}_q$.

A network code for $(\mathcal{G}, \mathcal{C})$ is defined to be a *feasible network code* if it achieves the given set of demands at the sinks i.e., if the zero-interference and the invertibility conditions are satisfied.

*A. System Model for time-varying LEKs*

When the LEKs are time-varying, we can't express the input-output relation as in (2). Hence, first, we need to derive the input-output relation involving transfer matrices which are dependent on varying LEKs. Retaining the notations as already introduced, we only point out the changes in the system model here.

Without loss of generality, we assume that a link between a pair of nodes has a unit delay (if the link has any other non-zero integer delay, we could introduce an appropriate number of dummy nodes in between the pair of nodes which are then connected by links of unit delays). For a given DAG $\mathcal{G}$ with integer delay on its links, define the adjacency matrix of $\mathcal{G}$ at time $t$ as the $|E| \times |E|$ matrix $K^{(t)}$, whose elements are given by

$$[K^{(t)}]_{ij} = \begin{cases} \beta_{e_i,e_j}^{(t)} & \text{head}(e_i) = \text{tail}(e_j) \\ 0 & \text{otherwise} \end{cases}.$$

Let the entries of $\mu \times |E|$ matrix $A^{(t)}$, at time $t$, be given by

$$[A^{(t)}]_{ij} = \begin{cases} \alpha_{l,e_j}^{(t)} & x_i = X_{\text{tail}(e_j)l} \\ 0 & \text{otherwise} \end{cases}.$$

Also, let the entries of $\nu \times |E|$ matrix $B^{(t)}$, at time $t$, be given by

$$[B^{(t)}]_{ij} = \begin{cases} \epsilon_{e_j,l}^{(t)} & y_i = Y_{\text{head}(e_j)l} \\ 0 & \text{otherwise} \end{cases}.$$

Let the set of vectors denoted by $\underline{\varepsilon}^{(t_1,t_2)}$ be the denote the set of LEKs from time instant $t_1$ to time instant $t_2$ ($t_2 \geq t_1$), i.e.,

$$\underline{\varepsilon}^{(t_1,t_2)} = \{\underline{\varepsilon}^{(t_1)}, \underline{\varepsilon}^{(t_1+1)}, \ldots, \underline{\varepsilon}^{(t_2)}\}$$

where $\underline{\varepsilon}^{(t_i)}$ denotes the LEKs at time $t_i$. Since the LEKs are time varying, the network transfer matrix is given by

$$M(D,t)^T = \Big( A^{(t-1)} I \ D + A^{(t-2)} K^{(t-1)} D^2 + A^{(t-3)} K^{(t-2)} K^{(t-1)} D^3$$
$$+ \ldots + A^{(t-d_{max})} K^{(t-(d_{max}-1))} .. K^{(t-2)} K^{(t-1)} D^{d_{max}} \Big) B^{(t)T}$$
$$\triangleq \sum_{d=0}^{d_{max}} M^{(d)T}(\underline{\varepsilon}^{(t-d,t)}) D^d$$

where $M^{(0)^T} = \mathbf{0}$, i.e., the zero matrix, as each link in the network is assumed to have a unit delay.

Since acyclic networks with delay are analogous to multiple transmitter-multiple receiver linear channel with time-varying impulse response between every transmitter and every receiver, the output symbols for the acyclic network with delay, at time instant $t$, at sink-$j$, is given by

$$\underline{Y_j}^{(t)} = \sum_{i=1}^{s} \sum_{d=0}^{d_{max}} M_{ij}^{(d)}(\underline{\varepsilon}^{(t-d,t)})\underline{X_i}^{(t-d)}. \quad (7)$$

## III. TRANSFORM TECHNIQUES FOR ACYCLIC NETWORKS WITH DELAY

In this section, we show that the output symbols at all the sinks which was originally a $\mathbb{F}_q$-linear combination of the input symbols across the different generations, at any given time instant, can be transformed into a $\mathbb{F}_q$-linear combination of the input symbols across the same generation.

Consider a matrix $A$ of size $n\nu \times n\mu$ given by

$$\begin{bmatrix} A_0 & A_1 & \cdots & A_{L-1} & A_L & 0 & 0 & \cdots & 0 \\ 0 & A_0 & \cdots & A_{L-2} & A_{L-1} & A_L & 0 & \cdots & 0 \\ \vdots & \vdots & \vdots & \vdots & \vdots & \vdots & \vdots & \vdots & \vdots \\ A_1 & A_2 & \cdots & A_L & 0 & 0 & 0 & \cdots & A_0 \end{bmatrix}$$

where, $A_i$s ($0 \leq i \leq L$) are matrices of size $\nu \times \mu$, whose elements belong to $\mathbb{F}_q$ and $n \gg L$. Note that the $(i+1)^{\text{th}}$ row of matrices is a circular shift of the $i^{\text{th}}$ row of matrices in $A$. We assume that $n$ divides $q-1$. Since $q = p^m$, $p$ and $n$ are coprime. The choice of $n$ is such that, there exists an $\alpha \in \mathbb{F}_q$ such that $n$ is the smallest integer for which $\alpha^n = 1$. This is indeed possible [9]. Define matrices $\hat{A}_j$ ($0 \leq j \leq n-1$), of size $\nu \times \mu$, as

$$\hat{A}_j = \sum_{i=0}^{L} \alpha^{(n-1-j)i} A_i.$$

Let $F$ be the finite-field DFT matrix given by

$$F = \begin{bmatrix} 1 & 1 & 1 & \cdots & 1 \\ 1 & \alpha & \alpha^2 & \cdots & \alpha^{n-1} \\ 1 & \alpha^2 & \alpha^4 & \cdots & \alpha^{2(n-1)} \\ \vdots & \vdots & \vdots & \vdots & \vdots \\ 1 & \alpha^{n-1} & \alpha^{2(n-1)} & \cdots & \alpha^{(n-1)(n-1)} \end{bmatrix}.$$

Also, define the matrix $Q_\mu$ as

$$Q_\mu = \begin{bmatrix} I_\mu & I_\mu & I_\mu & \cdots & I_\mu \\ I_\mu & \alpha I_\mu & \alpha^2 I_\mu & \cdots & \alpha^{n-1} I_\mu \\ I_\mu & \alpha^2 I_\mu & \alpha^4 I_\mu & \cdots & \alpha^{2(n-1)} I_\mu \\ \vdots & \vdots & \vdots & \vdots & \vdots \\ I_\mu & \alpha^{n-1} I_\mu & \alpha^{2(n-1)} I_\mu & \cdots & \alpha^{(n-1)(n-1)} I_\mu \end{bmatrix}. \quad (8)$$

Similarly we can define matrix $Q_\nu$. The following theorem will be useful in establishing the results subsequently.

*Theorem 1:* The matrix $A$ can be block diagonalized as

$$A = Q_\nu \hat{A} Q_\mu^{-1},$$

where, $\hat{A}$ is given by

$$\hat{A} = \begin{bmatrix} \hat{A}_{n-1} & 0 & 0 & \cdots & 0 \\ 0 & \hat{A}_{n-2} & 0 & \cdots & 0 \\ \vdots & \vdots & \vdots & \cdots & \vdots \\ 0 & 0 & \cdots & 0 & \hat{A}_0 \end{bmatrix}.$$

*Proof:* Proof is given in Appendix A. ∎

Now, consider an arbitrary acyclic network with delay. From (2) and (3),

$$\underline{Y_j}(D) = \sum_{i=1}^{s} M_{ij}(D)\underline{X_i}(D). \quad (9)$$

Now, consider a transmission scheme, where we take $n$ ($\gg d_{max}$) generations of input symbols at each source and first transmit last $d_{max}$ generations (which we call the *cyclic prefix*) followed by the $n$ generations of input symbols. Hence, $n + d_{max}$ time slots at each source are used for transmitting $n$ generations. Then, (9) can be written as (10) using (4). Now, after discarding first $d_{max}$ outputs at sink $j$, (10) can be re-written as (11) (given at the top of the next page). Using Theorem 1, (11) can be re-written as

$$\underline{Y_j}^n = \sum_{i=1}^{s} Q_{\nu_j} \hat{M}_{ij} Q_{\mu_i}^{-1} \underline{X_i}^n \quad (12)$$

where,

$$\underline{Y_j}^n = \begin{bmatrix} \underline{Y_j}^{(n-1)} \\ \underline{Y_j}^{(n-2)} \\ \vdots \\ \underline{Y_j}^{(0)} \end{bmatrix}; \quad \underline{X_i}^n = \begin{bmatrix} \underline{X_i}^{(n-1)} \\ \underline{X_i}^{(n-2)} \\ \vdots \\ \underline{X_i}^{(0)} \end{bmatrix};$$

$$\hat{M}_{ij} = \begin{bmatrix} \hat{M}_{ij}^{(n-1)} & 0 & 0 & \cdots & 0 \\ 0 & \hat{M}_{ij}^{(n-2)} & 0 & \cdots & 0 \\ \vdots & \vdots & \vdots & \cdots & \vdots \\ 0 & 0 & 0 & \cdots & \hat{M}_{ij}^{(0)} \end{bmatrix}.$$

Now, at each source $i$, transmit $\underline{X_i'}^n = Q_{\mu_i}\underline{X_i}^n$ instead of $\underline{X_i}^n$. Then, at each sink $j$, we receive $\underline{Y_j'}^n$. Let $\underline{Y_j}^n = Q_{\nu_j}^{-1}\underline{Y_j'}^n$. Then, from (12),

$$\underline{Y_j'}^n = \sum_{i=1}^{s} Q_{\nu_j} \hat{M}_{ij} Q_{\mu_i}^{-1} \underline{X_i'}^n$$

$$\underline{Y_j}^n = Q_{\nu_j}^{-1} \sum_{i=1}^{s} Q_{\nu_j} \hat{M}_{ij} Q_{\mu_i}^{-1} Q_{\mu_i} \underline{X_i}^n$$

$$\underline{Y_j}^n = \sum_{i=1}^{s} \hat{M}_{ij} \underline{X_i}^n \quad (13)$$

Now, (13) can be re-written as (for $0 \leq t \leq n-1$)

$$\underline{Y_j}^{(t)} = \sum_{i=1}^{s} \hat{M}_{ij}^{(t)} \underline{X_i}^{(t)}. \quad (14)$$

Hence, each element of $\underline{Y_j}^{(t)}$ is a $\mathbb{F}_q$-linear combination of the input symbols across the same generation. We now say

$$\begin{bmatrix} \underline{Y_j}^{(n-1)} \\ \underline{Y_j}^{(n-2)} \\ \vdots \\ \underline{Y_j}^{(0)} \\ \underline{Y_j}^{(-1)} \\ \vdots \\ \underline{Y_j}^{(-d_{max})} \end{bmatrix} = \sum_{i=1}^{3} \begin{bmatrix} M_{ij}^{(0)} & M_{ij}^{(1)} & \cdots & M_{ij}^{(d_{max})} & 0 & 0 & \cdots & 0 & 0 \\ 0 & M_{ij}^{(0)} & \cdots & M_{ij}^{(d_{max}-1)} & M_{ij}^{(d_{max})} & 0 & \cdots & 0 & 0 \\ \vdots & \vdots & \vdots & \vdots & \ddots & \ddots & \ddots & \vdots & \vdots \\ 0 & 0 & \cdots & 0 & M_{ij}^{(0)} & M_{ij}^{(1)} & \cdots & M_{ij}^{(d_{max}-1)} & M_{ij}^{(d_{max})} \\ 0 & 0 & \cdots & 0 & 0 & M_{ij}^{(0)} & \cdots & M_{ij}^{(d_{max}-2)} & M_{ij}^{(d_{max}-1)} \\ \vdots & \vdots & \vdots & \vdots & \vdots & \vdots & \vdots & \vdots & \vdots \\ 0 & 0 & \cdots & 0 & 0 & 0 & 0 & 0 & M_{ij}^{(0)} \end{bmatrix} \begin{bmatrix} \underline{X_i}^{(n-1)} \\ \underline{X_i}^{(n-2)} \\ \vdots \\ \underline{X_i}^{(0)} \\ \underline{X_i}^{(n-1)} \\ \vdots \\ \underline{X_i}^{(n-d_{max})} \end{bmatrix} \quad (10)$$

$$\begin{bmatrix} \underline{Y_j}^{(n-1)} \\ \underline{Y_j}^{(n-2)} \\ \vdots \\ \underline{Y_j}^{(0)} \end{bmatrix} = \sum_{i=1}^{s} \underbrace{\begin{bmatrix} M_{ij}^{(0)} & M_{ij}^{(1)} & \cdots & M_{ij}^{(d_{max}-1)} & M_{ij}^{(d_{max})} & 0 & \cdots & 0 & 0 & 0 \\ 0 & M_{ij}^{(0)} & \cdots & M_{ij}^{(d_{max}-2)} & M_{ij}^{(d_{max}-1)} & M_{ij}^{(d_{max})} & \cdots & 0 & 0 & 0 \\ \vdots & \vdots & \vdots & \vdots & \vdots & \vdots & \vdots & \vdots & \vdots & \vdots \\ M_{ij}^{(1)} & M_{ij}^{(2)} & \cdots & M_{ij}^{(d_{max})} & 0 & 0 & \cdots & 0 & 0 & M_{ij}^{(0)} \end{bmatrix}}_{M_{ij}} \begin{bmatrix} \underline{X_i}^{(n-1)} \\ \underline{X_i}^{(n-2)} \\ \vdots \\ \underline{X_i}^{(0)} \end{bmatrix} \quad (11)$$

that we have transformed the ayclic network with delay into $n$-*instantaneous networks*.

*Remark 1:* Note that the linear processing of multiplying by matrices $Q_{\mu_i}$ at source-$i$ and $Q_{\nu_j}^{-1}$ at sink-$j$ are done in a distributed fashion which is necessary because the sources and sinks are distributed in the actual network.

*Remark 2:* One can observe that transmitting $\underline{X_i'}^n = Q_{\mu_i}\underline{X_i}^n$ implies taking DFT across $n$ generations of each of the $\mu_i$ random-processes generated at source-$i$. Similarly, the pre-multiplication by $Q_{\nu_j}^{-1}$ at sink-$j$ simply implies taking IDFT across $n$ generations of each of the $\nu_j$ random-processes received. The entire processing, including addition of cyclic prefix at source-$i$ and removal of cyclic prefix at sink-$j$ is shown in a block diagram in Fig. 1 (given at the top of the next page).

Now, let us re-write (14) as

$$\underline{Y_j}^{(t)} = \sum_{i=1}^{s} \sum_{l_i=1}^{\mu_i} \hat{M}_{ij}^{(t)}(l_i)\underline{X_i}^{(t)}(l_i).$$

where $\hat{M}_{ij}^{(t)}(l_i)$ denotes the $l_i^{\text{th}}$ column of $\hat{M}_{ij}^{(t)}$ and $\underline{X_i}^{(t)}(l_i)$ denotes the $l_i^{\text{th}}$ element of $\underline{X_i}^{(t)}$.

Similar to the zero-interference and invertibility conditions in Lemma 1, we have the following theorem for solvability of (14).

*Theorem 2:* An acylic network $(\mathcal{G}, \mathcal{C})$ with delay, incorporating the transform techniques, is solvable iff there exists an assignment to variables $\underline{\varepsilon}$ such that:

1) *Zero-Interference:* $\hat{M}_{ij}^{(t)}(l_i) = 0$ for all pairs (source-$i$, sink-$j$) of nodes such that (source-$i$, sink-$j$, $\underline{X_i}^{(t)}(l_i)$) $\notin \mathcal{C}_j$ for $0 \leq t \leq n-1$.
2) *Invertibility:* If $\mathcal{C}_j$ contains the connections (source-$i_1$, sink-$j$, $\underline{X_{i_1}}^{(t)}(l_{i_1})$), (source-$i_2$, sink-$j$, $\underline{X_{i_2}}^{(k)}(l_{i_2})$), $\cdots$, (source-$i_{s'}$, sink-$j$, $\underline{X_{i_{s'}}}^{(t)}(l_{i_{s'}})$), then, the sub-matrix $[\hat{M}_{i_1j}^{(t)}(l_{i_1}) \ \hat{M}_{i_2j}^{(t)}(l_{i_2}) \ \cdots \ \hat{M}_{i_{s'}j}^{(t)}(l_{i_{s'}})]$ is a nonsingular $\nu_j \times \nu_j$ matrix for $0 \leq t \leq n-1$.

The network code which satisfies the invertibility and the zero-interference conditions for $(\mathcal{G}, \mathcal{C})$ in the transform approach using a suitable choice of $\alpha$ for the DFT operations is defined as an *feasible transform network code* for $(\mathcal{G}, \mathcal{C})$.

### A. Existence of a network code in the transform approach

In this section, we prove that under certain conditions there exists a feasible network code for a given $(\mathcal{G}, \mathcal{C})$ if and only if there exists a feasible transform network code. Towards that end, we prove a lemma. We first define the polynomial $f(D)$ which will be used henceforth throughout this paper.

$$f(D) = \prod_{j=1}^{r} det\left(M_j'(D)\right). \quad (15)$$

where, $M_j'(D)$ is the square submatrix of $M_j(D)$ indicating the source processes that are demanded by sink-$j$.

*Lemma 2:* Suppose there exists a feasible network code for $(\mathcal{G}, \mathcal{C})$ over some field $\mathbb{F}_q$. For some $\alpha \in \mathbb{F}_{q^a}$ (for some positive integer $a$), the local encoding kernels defined by the feasible network code for $(\mathcal{G}, \mathcal{C})$ (viewed in the extension field $\mathbb{F}_{q^a}$) along with the DFT operations defined using $\alpha$ result in a feasible transform network code for $(\mathcal{G}, \mathcal{C})$ if and only if $f(\alpha^t) \neq 0$ for all $0 \leq t \leq n-1$.

*Proof:* Proof is given in Appendix C. ∎

We now prove the following theorem which concerns with the relationship between the existence of a feasible network code and a feasible transform network code for $(\mathcal{G}, \mathcal{C})$.

*Theorem 3:* Let $(\mathcal{G}, \mathcal{C})$ be the given acyclic delay network with the set of connections $\mathcal{C}$ demanded by the sinks. There exists a feasible transform network code for $(\mathcal{G}, \mathcal{C})$ if and only if there exists a feasible network code for $(\mathcal{G}, \mathcal{C})$ such that $(D-1) \nmid f(D)$.

*Proof:* Proof is given in Appendix D. ∎

*Remark 3:* Based on the constructive proof of Theorem 3, a large field might be required for the existence of a suitable value for $\alpha$ that defines the necessary transform for the network, under the condition that the rate-loss $\left(\frac{d_{max}}{n}\right)$ due to the transform approach be less. The transformed network would then have to be operated over this large field, i.e., the matrices $\hat{M}_{ij}^{(t)}$ have elements from this large field (which is at least a degree $n$ extension over the base field over which the non-transform network code is defined). It is known that (see [11], for example) inverting a $\nu_j \times \nu_j$ matrix (at some sink-$j$) takes $O(\nu_j^3)$ computations, however over the extension field. In the process of computing these inverses, the information symbols corresponding to the $n$ generations are obtained by Gauss-Jordan elimination. In terms of base field computational complexity, the complexity of computing the inverse of the transfer matrix becomes $O\left(\nu_j^3 n(\log n)(\log \log n)\right)$, as each multiplication in the extension field involves $O\left(n(\log n)(\log \log n)\right)$ computations over the base field [12] (it is equivalent to multiplying two polynomials of degree at least $n-1$ over the base field). The total complexity of recovering the input symbols at all the $n$ generations is then $O\left(n^2 \nu_j^3 (\log n)(\log \log n)\right)$.

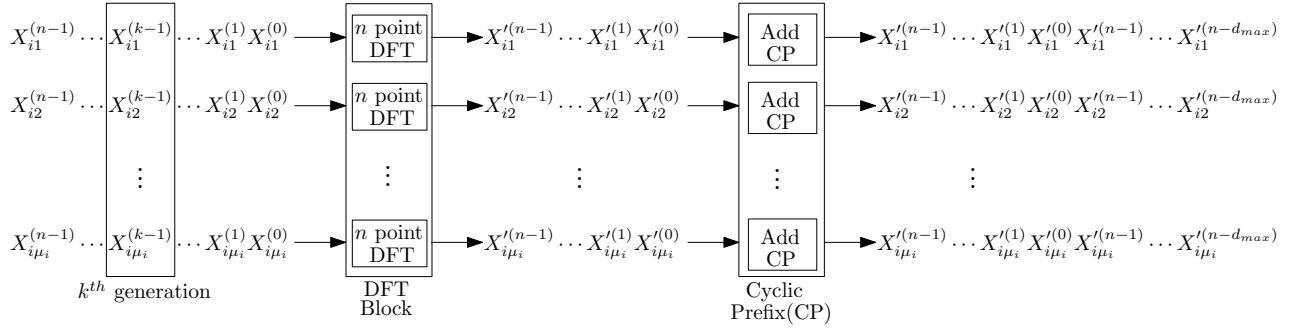

(a) Linear Processing at Source-$i$

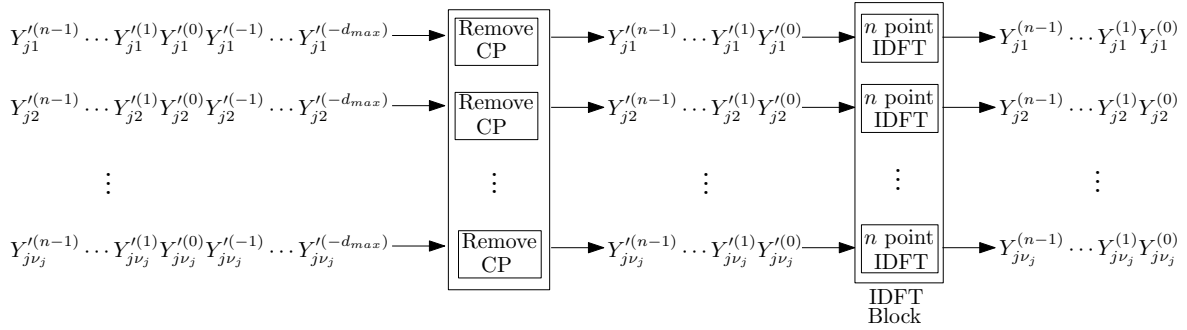

(b) Linear Processing at Sink-$j$

Fig. 1. Block Diagram to illustrate linear processing at Source-$i$ and Sink-$j$.

On the other hand, if the non-transform network code is used as such, the transfer matrices $M_j'(D)$ consist of polynomials of degree upto $d_{max}$ in $D$ over the base field. Again, it is known (see [11], for example) that finding the inverse of such a matrix has complexity $O(\nu_j^3 d_{max})$. To do a fair comparison with the transform case, we consider decoding of $n$-generations ($n$ being large as in the transform case) of information. Note that inversion of the matrix $M_j'(D)$ does not give us the information polynomials directly. A naive method of obtaining the each information polynomial would then require $\nu_j^2$ multiplications of polynomials over the base field (each of which has complexity $O\left(n(\log n)(\log \log n)\right)$, assuming that $\nu_j d_{max} < n$.) and one division between polyno-mials (again with complexity $O\left(n(\log n)(\log \log n)\right)$). Therefore, the total complexity involved in recovering the information sequences would then be $O\left(\nu_j^3 n(\log n)(\log \log n)\right) + O\left(\nu_j n(\log n)(\log \log n)\right) + O(\nu_j^3 d_{max})$ computations.

Thus, we see that there is an advantage in the complexity of decoding in the non-transform network compared to the transform network (inspite of using the least possible size for the extension field). Therefore, complexity reduction is not an advantage of the transform process.

We now present an example acyclic network in which there exists a feasible network code, using which we obtain a feasible transform network code for some choice of $n \geq 7$.

*Example 1:* Consider the network $\mathcal{G}$ shown in Fig. 2. This is a unit-delay network (where each edges have a delay of one unit associated with it) taken from [13]. For $1 \leq i \leq 3$, each source $s_i$ has an information sequence $x_i(D)$. This network has non-multicast demands, with sinks $u_j : 1 \leq j \leq 3$ requiring all three information sequences, while sink $u_4$ requires $\{x_1(D), x_3(D)\}$ and $u_5$ demands $\{x_2(D), x_3(D)\}$. Let $\mathcal{C}$ denote these set of demands. A feasible network code for $(\mathcal{G}, \mathcal{C})$ over $\mathbb{F}_2$ as obtained in [13] can be obtained by using $1$ as the local encoding kernel coefficient at all non-sink nodes. The transfer matrix $M_{u_j}(D)$, the invertible submatrix $M_{u_j}'(D)$ of $M_{u_j}(D)$, and their determinants for the sinks $u_j : 1 \leq j \leq 5$ are tabulated in Table 1.

We therefore have $f(D) = D^{25}$. Note that $f(1) \neq 0$ and $d_{max} = 4$ for this network. Therefore, with $n = 2^m - 1$ for any positive integer $m \geq 3$, i.e., $\alpha$ being the primitive element of $\mathbb{F}_{2^m}$, we will then have $f(\alpha^t) \neq 0$ for any $0 \leq t \leq n-1$.

TABLE I

| Sink | Network transfer matrix $M_{u_j}(D)$ | Invertible submatrix $M'_{u_j}(D)$ of $M_{u_j}(D)$ | Determinant of $M'_{u_j}(D)$, $det(M'_{u_j}(D))$ |
|---|---|---|---|
| $u_1$ | $\begin{pmatrix} D & 0 & 0 \\ 0 & D & 0 \\ D^3 & D^3 & D^3 \end{pmatrix}$ | $M_{u_1}(D)$ | $D^5$ |
| $u_2$ | $\begin{pmatrix} D & 0 & 0 \\ 0 & 0 & D \\ D^3 & D^3 & D^3 \end{pmatrix}$ | $M_{u_2}(D)$ | $D^5$ |
| $u_3$ | $\begin{pmatrix} 0 & D & 0 \\ 0 & 0 & D^2 \\ D^3 & D^3 & D^3 \end{pmatrix}$ | $M_{u_3}(D)$ | $D^6$ |
| $u_4$ | $\begin{pmatrix} D^4 & 0 & D^4+D^5 \\ 0 & 0 & D \end{pmatrix}$ | $\begin{pmatrix} D^4 & D^4+D^5 \\ 0 & D \end{pmatrix}$ | $D^5$ |
| $u_5$ | $\begin{pmatrix} 0 & D^3 & D^4 \\ 0 & 0 & D \end{pmatrix}$ | $\begin{pmatrix} D^3 & D^4 \\ 0 & D \end{pmatrix}$ | $D^4$ |

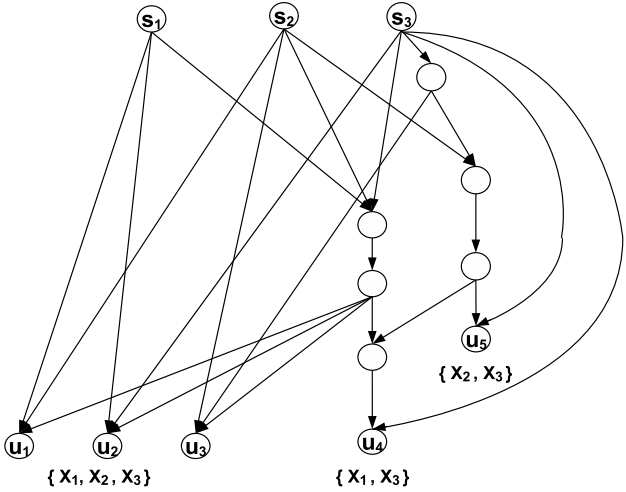

Fig. 2. A unit-delay network with 3 sources and 5 sinks

By Lemma 2, we then have a feasible transform network code for $(\mathcal{G}, \mathcal{C})$.

In the next section we shall apply these transform techniques to three-source three-destination unicast network with delays.

## IV. THREE SOURCE-THREE DESTINATION UNICAST NETWORK WITH DELAYS

In [6], the concept of *interference alignment* from interference channels [7], was extended to instantaneous acyclic unicast networks with three source-destination pairs for the case where, each source-destination pair has a min-cut of 1 and where, the zero-interference conditions in Theorem 6 of [3] cannot be satisfied. This was called *network alignment*. It was shown, in [6], that for a class of such networks, it is possible to achieve a throughput close to $1/2$ for every source-destination pair via network alignment.

In this section, we deal with acyclic three source-three destination unicast network with delays, with each source-destination pair having a min-cut of 1. We employ the results from the previous section and show that, even when the zero-interference conditions of Lemma 1 cannot be satisfied, for a class of three source-three destination unicast networks with arbitrary integer delays on its links, we can achieve a throughput close to $1/2$ for every source-destination pair by making use of network alignment. We take two approaches in achieving this - using time-invariant LEKs and using time-varying LEKs.

Let the random process injected into the network by source $S_i$ ($i \in (1,2,3)$) be $X_i(D)$. Source $S_i$ needs to communicate only with destination $D_i$ ($i \in (1,2,3)$). Here, $\mu_i = 1$ and $\nu_j = 1$ ($(i,j) \in 1,2,3$).

We shall consider the following two cases separately.

1) The min-cut between source-$i$ and sink-$j$ is greater than or equal to 1, for all $i \neq j$.
2) The min-cut between source-$i$ and sink-$j$ is equal to 0, for some $i \neq j$.

**Case 1:** The min-cut between source-$i$ and sink-$j$ is greater than or equal to 1, for all $i \neq j$.

### A. Achieving a Throughput of $1/2$ with Time-Invariant LEKs

Now, consider a transmission scheme, where we take $2n+1$ ($\gg d_{max}$) generations of input symbols at each source and first transmit the last $d_{max}$ generations (i.e., the cyclic prefix) followed by the $2n+1$ generations of input symbols. Let $Q_1 X_i^{2n+1}$ be the input symbols transmitted by source $i$, where,

$$X_i^{2n+1} = [X_i^{(2n)}\ X_i^{(2n-1)}\ \cdots\ X_i^{(0)}]^T$$

Also, let $X_1^{2n+1} = V_1 X_1'^{n+1}$, $X_2^{2n+1} = V_2 X_2'^n$, and $X_3^{2n+1} = V_3 X_3'^n$, where, $V_1$ is a $(2n+1) \times (n+1)$ matrix,

$V_2$ is a $(2n+1) \times n$ matrix, $V_3$ is a $(2n+1) \times n$ matrix, and

$$X_1'^{n+1} = [X_1'^{(0)} \ X_1'^{(1)} \ \cdots \ X_1'^{(n)}]^T$$
$$X_2'^{n} = [X_2'^{(0)} \ X_2'^{(1)} \ \cdots \ X_2'^{(n-1)}]^T$$
$$X_3'^{n} = [X_3'^{(0)} \ X_3'^{(1)} \ \cdots \ X_3'^{(n-1)}]^T.$$

The quantities $X_1'^{n+1}$, $X_2'^{n}$ and $X_3'^{n}$ denote the $(n+1)$, $n$, and $n$ independent input symbols generated by sources-1, 2 and 3 respectively. Now, from (13), for $j \in \{1,2,3\}$,

$$Y_j^{2n+1} = \hat{M}_{1j} V_1 X_1'^{n+1} + \hat{M}_{2j} V_2 X_2'^{n} + \hat{M}_{3j} V_3 X_3'^{n},$$

where, $Y_j^{2n+1}$ denotes the $(2n+1)$ output symbols at sink-$j$. The objective is to recover the $(n+1)$ independent input symbols of source-1, $n$ independent input symbols of source-2 and $n$ independent input symbols of source-3 at sinks-1, 2 and 3 from $Y_1^{2n+1}$, $Y_2^{2n+1}$ and $Y_3^{2n+1}$ respectively.

For acyclic networks without delay, the network alignment concept in [6] involved varying LEKs at every time instant. But with delays it is possible, in some cases, to achieve network alignment even with time-invariant LEKs. This is what we show in this sub-section.

First, note that the elements of $\hat{M}_{ij}$s are functions of $\underline{\varepsilon}$.

*Lemma 3:* Determinant of the matrix $\hat{M}_{ij}$ $\forall$ $(i,j) \in \{1,2,3\}$ is a non-zero polynomial in $\underline{\varepsilon}$.

*Proof:* Proof is given in Appendix E ∎

Let

$$a^{(k)} = \hat{M}_{21}^{(k)} \hat{M}_{32}^{(k)} \hat{M}_{13}^{(k)} \ (k \in \{0, 1, .., 2n\})$$
$$b^{(k)} = \hat{M}_{31}^{(k)} \hat{M}_{23}^{(k)} \hat{M}_{12}^{(k)} \ (k \in \{0, 1, .., 2n\})$$
$$T = \hat{M}_{21} \hat{M}_{32} \hat{M}_{13} \hat{M}_{31}^{-1} \hat{M}_{23}^{-1} \hat{M}_{12}^{-1} \quad (16)$$
$$R = \hat{M}_{13} \hat{M}_{23}^{-1} \quad (17)$$
$$S = \hat{M}_{12} \hat{M}_{32}^{-1}. \quad (18)$$

Now, choose

$$V_1 = [W \ TW \ T^2 W \ \cdots \ T^n W] \quad (19)$$
$$V_2 = [RW \ RTW \ RT^2 W \ \cdots \ RT^{n-1} W] \quad (20)$$
$$V_3 = [STW \ ST^2 W \ \cdots \ ST^n W] \quad (21)$$

where, $W = [1 \ 1 \ \cdots \ 1]^T$ (all ones vector of size $(2n+1) \times 1$). Since the transform approach requires that $2n+1 | p^m - 1$, we shall find it useful in stating the exact relationship between $2n+1$ and $p$ which will be used in the result that follows.

*Lemma 4:* The positive integer $2n+1$ divides $p^m - 1$ for some positive integer $m$ iff $p \nmid 2n+1$.

*Proof:* Proof is given in Appendix F. ∎

*Theorem 4:* The input symbols $X_1'^{n+1}$, $X_2'^{n}$, and $X_3'^{n}$ can be exactly recovered at $T_1$, $T_2$, and $T_3$ from the output symbols $Y_1^{2n+1}$, $Y_2^{2n+1}$, and $Y_3^{2n+1}$ respectively subject to $p \nmid 2n+1$, if the following conditions hold.

$$\text{Rank}[V_1 \ \hat{M}_{11}^{-1} \hat{M}_{21} V_2] = 2n+1 \quad (22)$$
$$\text{Rank}[\hat{M}_{12}^{-1} \hat{M}_{22} V_2 \ V_1] = 2n+1 \quad (23)$$
$$\text{Rank}[\hat{M}_{13}^{-1} \hat{M}_{33} V_3 \ V_1] = 2n+1 \quad (24)$$

*Proof:* Proof is given in Appendix G ∎

When the conditions of the above Theorem are satisfied, we say that network alignment is feasible. When network alignment is feasible, throughputs of $\frac{(n+1)}{(2n+1)}$, $\frac{n}{(2n+1)}$, and $\frac{n}{(2n+1)}$ are achieved for the source-destination pairs $S_1 - D_1$, $S_2 - D_2$, and $S_3 - D_3$ respectively. Hence, as $n \to \infty$, a throughput of $1/2$ is achieved for every source-destination pair.

*Remark 4:* To satisfy (22)-(24), we have to first ensure that $V_1$ is full-rank. Note that

$$T = \begin{bmatrix} \frac{a^{(2n)}}{b^{(2n)}} & 0 & \cdots & 0 \\ 0 & \frac{a^{(2n-1)}}{b^{(2n-1)}} & \cdots & 0 \\ \vdots & \vdots & \vdots & \vdots \\ 0 & 0 & \cdots & \frac{a^{(0)}}{b^{(0)}} \end{bmatrix}.$$

Now, any collection of $(n+1)$ rows of $V_1$ is a Vandermonde matrix whose determinant is a non-zero polynomial iff $\frac{a^{(k_1)}}{b^{(k_1)}} \neq \frac{a^{(k_2)}}{b^{(k_2)}}$ is satisfied for every $k_1, k_2 \in \{0, 1, .., 2n\}$ and $k_1 \neq k_2$. But, for all the columns of $V_1$ to be independent, it is enough if there exists atleast $(n+1)$ linearly independent rows. This condition is satisfied if there are atleast $(n+1)$ distinct $\frac{a^{(k_1)}}{b^{(k_1)}}$s ($k_1 \in \{0, 1, .., 2n\}$). If $V_1$ is full-rank, then, by the choice of $V_2$ and $V_3$ in (20) and (21) respectively, $\hat{M}_{11}^{-1} \hat{M}_{21} V_2$, $\hat{M}_{12}^{-1} \hat{M}_{22} V_2$ and $\hat{M}_{13}^{-1} \hat{M}_{33} V_3$ are also full-rank matrices. Hence, when there are atleast $(n+1)$ distinct $\frac{a^{(k_1)}}{b^{(k_1)}}$s ($k_1 \in \{0, 1, .., 2n\}$), the choice of $V_1$, $V_2$ and $V_3$ atleast ensures that $V_1$, $V_2$ and $V_3$ are individually full-rank matrices.

*Remark 5:* Note that, for three-source three-destination unicast network without delay, considered in [6], it was not possible to achieve network alignment without changing the LEKs with time. When there is no delay, the matrices $T$, $R$, and $S$, given in (16)-(18), would simply be equal to $f(\underline{\varepsilon}) I_{2n+1}$ (where, $f(\underline{\varepsilon})$ is some polynomial in $\underline{\varepsilon}$) and hence, the matrices $V_1$, $V_2$ and $V_3$ as given in (19)-(21) are themselves not full-rank matrices. Hence, $\underline{\varepsilon}$ was varied with time in [6]. However, in the case of delay it is easy to see from the structure of the matrix $\hat{M}_{ij}$ that the matrices $T$, $R$, and $S$ are not necessarily scaled identity matrices.

The following example, taken from [6] (but with delays), illustrates the existence of a network where network alignment is feasible with time-invariant LEKs.

*Example 2:* Consider the network shown in Fig. 3 (at the top of the next page). Each link is taken to have unit-delay. In accordance with the LEKs denoted as in the figure, the transfer matrices $M_{ij}(D)$ are as given below.

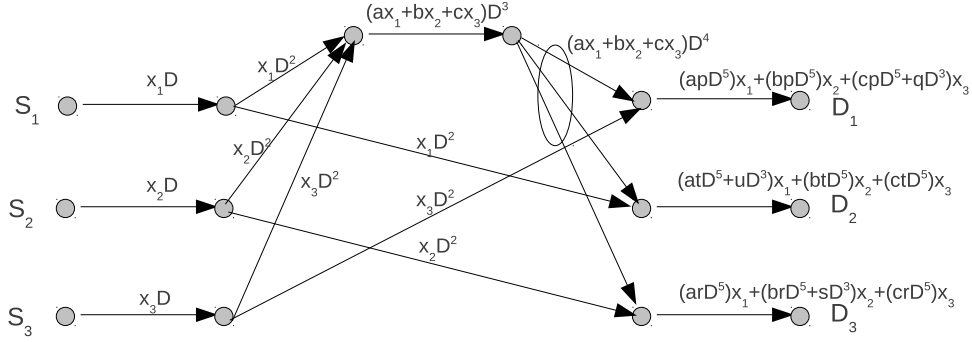

Fig. 3. A three-source three-destination unicast network where network alignment with time-invariant LEKs is feasible.

$$M_{11}(D) = M_{11}^{(5)}D^5 = apD^5,$$
$$M_{12}(D) = M_{12}^{(3)}D^3 + M_{12}^{(5)}D^5 = uD^3 + atD^5,$$
$$M_{13}(D) = M_{13}^{(5)}D^5 = arD^5$$
$$M_{21}(D) = M_{21}^{(5)}D^5 = bpD^5,$$
$$M_{22}(D) = M_{22}^{(5)}D^5 = btD^5$$
$$M_{23}(D) = M_{23}^{(3)}D^3 + M_{23}^{(5)}D^5 = sD^3 + brD^5$$
$$M_{31}(D) = M_{31}^{(3)}D^3 + M_{31}^{(5)}D^5 = qD^3 + cpD^5$$
$$M_{32}(D) = M_{32}^{(5)}D^5 = ctD^5,$$
$$M_{33}(D) = M_{33}^{(5)}D^5 = crD^5.$$

Note that the network does not satisfy the zero-interference conditions of Lemma 1. Here, $d_{max} = 2$. It can be verified that network alignment is feasible with $2n + 1 = 7$. Specifically, network alignment works with the following assignment to the LEKs.

$$a = b = c = p = r = t = 1$$
$$s = 1 + \beta^2 + \beta^3 + \beta^4 + \beta^5$$
$$q = 1 + \beta + \beta^2$$
$$u = 1 + \beta^4$$
$$\alpha = \beta^9$$

where, $\beta$ is the primitive element of $GF(2^6)$, i.e., root of the minimal polynomial $(1 + x + x^6)$.

### B. Achieving a Throughput of $1/2$ with Time-Varying LEKs

In this section, we shall generalize the selection of matrices $V_1$, $V_2$ and $V_3$ and hence Theorem 4 along with the use of time-varying LEKs. Throughout the sub-section we shall assume that the LEKs and the other variables that we shall encounter belong to the algebraic closure of the field $\mathbb{F}_p$ which is denoted by $\overline{\mathbb{F}_p}$. Clearly, once an assignment to the LEKs and variables are made, they belong to a finite extension of $\mathbb{F}_p$.

In this case of time-varying LEKs, the network cannot be decomposed into $(2n + 1)$ instantaneous networks using the transform method. This is explained below.

Consider a transmission scheme, where we take $2n+1 (\gg d_{max})$ generations of input symbols at each source and first transmit last $d_{max}$ generations (i.e., the cyclic prefix) followed by the $2n+1$ generations of input symbols. Let $X_i^{2n+1}$ be the input symbol transmitted by source-$i$, where,

$$X_i^{2n+1} = [X_i^{(2n)} \ X_i^{(2n-1)} \ \cdots \ X_i^{(0)}]^T.$$

Also, let $X_1^{2n+1} = V_1 X_1'^{n+1}$, $X_2^{2n+1} = V_2 X_2'^n$, and $X_3^{2n+1} = V_3 X_3'^n$, where, $V_1$ is a $(2n+1) \times (n+1)$ matrix, $V_2$ is a $(2n+1) \times n$ matrix, $V_3$ is a $(2n+1) \times n$ matrix, and

$$X_1'^{n+1} = [X_1'^{(0)} \ X_1'^{(1)} \ \cdots \ X_1'^{(n)}]^T$$
$$X_2'^n = [X_2'^{(0)} \ X_2'^{(1)} \ \cdots \ X_2'^{(n-1)}]^T$$
$$X_3'^n = [X_3'^{(0)} \ X_3'^{(1)} \ \cdots \ X_3'^{(n-1)}]^T.$$

The quantities $X_1'^{n+1}$, $X_2'^n$ and $X_3'^n$ denote the $(n+1)$, $n$, and $n$ independent input symbols generated by sources-1, 2, and 3 respectively. Now, from (7) and following the same steps involved in writing (10) and (11), for $j \in \{1, 2, 3\}$, we get (25) (given at the top of the next page). In brief, for $j \in \{1, 2, 3\}$, we have

$$Y_j^{2n+1} = M_{1j} V_1 X_1'^{n+1} + M_{2j} V_2 X_2'^n + M_{3j} V_3 X_3'^n$$

where, $Y_j^{2n+1}$ denotes the $(2n+1)$ output symbols at sink-$j$ and $M_{ij}$ is as given in (25). The objective is to recover the $(n+1)$ independent input symbols of source-1, $n$ independent input symbols of source-2 and $n$ independent input symbols of source-3 at sinks-1, 2 and 3 from $Y_1^{2n+1}$, $Y_2^{2n+1}$ and $Y_3^{2n+1}$ respectively.

$$\begin{aligned}
&[Y_j{}^{(2n)}\ Y_j{}^{(2n-1)}\ \cdots\ Y_j{}^{(0)}]^T \\
&= \sum_{i=1}^{s} \\
&\underbrace{\begin{bmatrix}
M_{ij}^{(0)}(\underline{\varepsilon}^{(2n,2n)}) & M_{ij}^{(1)}(\underline{\varepsilon}^{(2n-1,2n)}) & \cdots & M_{ij}^{(d_{max}-1)}(\underline{\varepsilon}^{(2n-d_{max}+1,2n)}) & M_{ij}^{(d_{max})}(\underline{\varepsilon}^{(2n-d_{max},2n)}) & & & \\
0 & M_{ij}^{(0)}(\underline{\varepsilon}^{(2n-1,2n-1)}) & \cdots & M_{ij}^{(d_{max}-2)}(\underline{\varepsilon}^{(2n-d_{max}+1,2n-1)}) & M_{ij}^{(d_{max}-1)}(\underline{\varepsilon}^{(2n-d_{max},2n-1)}) & & & \\
\vdots & \vdots & \vdots & \vdots & \vdots & & & \\
M_{ij}^{(1)}(\underline{\varepsilon}^{(-1,0)}) & M_{ij}^{(2)}(\underline{\varepsilon}^{(-2,0)}) & \cdots & M_{ij}^{(d_{max})}(\underline{\varepsilon}^{(-d_{max},0)}) & 0 & & & \\
& & & & M_{ij}^{(d_{max})}(\underline{\varepsilon}^{(2n-1-d_{max},2n-1)}) & 0 & \cdots & 0 \\
& & & & \vdots & \vdots & \vdots & \vdots \\
& & & & 0 & 0 & \cdots & M_{ij}^{(0)}(\underline{\varepsilon}^{(0,0)})
\end{bmatrix}}_{M_{ij}} \\
&\times [X_i{}^{(2n)}\ X_i{}^{(2n-1)}\ \cdots\ X_i{}^{(0)}]^T
\end{aligned} \quad (25)$$

Note that $M_{ij}$ is not a circulant matrix and cannot be diagonalized in general. Let $\underline{\varepsilon}' = \{\underline{\varepsilon}^{(-d_{max})}, \underline{\varepsilon}^{(-d_{max}+1)}, ..., \underline{\varepsilon}^{(2n)}\}$.

*Lemma 5:* Determinant of the matrix $M_{ij}\ \forall\ (i,j) \in \{1,2,3\}$ is a non-zero polynomial in $\underline{\varepsilon}'$.

*Proof:* Proof is given in Appendix H ∎

Hence, the inverse of $M_{ij}$ exists. Now, let the elements of $V_1$ be given by

$$[V_1]_{ij} = \theta_{ij};\ i \in \{1,2,..,2n+1\}, j \in \{1,2,..,n+1\}, \quad (26)$$

where $\theta_{ij}$ is a variable that takes values from $\overline{\mathbb{F}_p}$. Also, let

$$V_2 = M_{23}^{-1}M_{13}V_1 A \text{ and } V_3 = M_{32}^{-1}M_{12}V_1 B \quad (27)$$

where, the elements of the matrices $A$ and $B$, of size $(n+1) \times n$, are given by $[A]_{ij} = a_{ij}$ and $[B]_{ij} = b_{ij}$ respectively ($a_{ij}$ and $b_{ij}$ are variables that take values from $\overline{\mathbb{F}_p}$). Let

$$T_1 = M_{12}^{-1}M_{32}M_{31}^{-1}M_{21}M_{23}^{-1}M_{13}.$$

Also, let $\underline{\theta} = \{\theta_{ij} | i \in \{1,2,..,2n+1\},\ j \in \{1,2,..,n+1\}\}$, $\underline{a} = \{a_{ij} | i \in \{1,2,..,n+1\},\ j \in \{1,2,..,n\}\}$ and $\underline{b} = \{b_{ij} | i \in \{1,2,..,n+1\},\ j \in \{1,2,..,n\}\}$. Let

$$\begin{aligned}
f_1(\underline{\theta}, \underline{\varepsilon}', \underline{a}) &= det([V_1\ \ M_{11}^{-1}M_{21}V_2]) \\
f_2(\underline{\theta}, \underline{\varepsilon}', \underline{a}) &= det([M_{12}^{-1}M_{22}V_2\ \ V_1]) \\
f_3(\underline{\theta}, \underline{\varepsilon}', \underline{b}) &= det([M_{13}^{-1}M_{33}V_3\ \ V_1]) \\
f_4(\underline{\varepsilon}') &= \prod_{(i,j)\in\{1,2,3\}} det(M_{ij}) \\
f(\underline{\theta}, \underline{\varepsilon}', \underline{a}, \underline{b}) &= f_1(\underline{\theta}, \underline{\varepsilon}', \underline{a}) f_2(\underline{\theta}, \underline{\varepsilon}', \underline{a}) f_3(\underline{\theta}, \underline{\varepsilon}', \underline{b}) f_4(\underline{\varepsilon}').
\end{aligned}$$

Denote the elements of a matrix $C$, of size $n \times n$, by $[C]_{ij} = c_{ij}$, where $c_{ij}$ is a variable that takes values from $\overline{\mathbb{F}_p}$ ($i \in \{1,2,...,n\}, j \in \{1,2,...,n\}$). Let $\underline{c} = \{c_{ij} | i \in \{1,2,..,n\},\ j \in \{1,2,..,n\}\}$. For $i \in \{1,2,..,2n+1\}$ and $j \in \{1,2,..,n\}$, let

$$g_{ij}(\underline{\theta}, \underline{\varepsilon}', \underline{a}, \underline{b}, \underline{c}) = [T_1 V_1 A]_{ij} - [V_1 B C]_{ij}.$$

Let $g_{ij}^{(nr)}(\underline{\theta}, \underline{\varepsilon}', \underline{a}, \underline{b}, \underline{c})$ and $g_{ij}^{(dr)}(\underline{\theta}, \underline{\varepsilon}', \underline{a}, \underline{b}, \underline{c})$ respectively denote the numerator and denominator of the rational-polynomial $g_{ij}(\underline{\theta}, \underline{\varepsilon}', \underline{a}, \underline{b}, \underline{c})$ ($i \in \{1,2,..,2n+1\}$ and $j \in \{1,2,..,n\}$).

Similar notation is used for the numerator and denominator of $f(\underline{\theta}, \underline{\varepsilon}', \underline{a}, \underline{b})$. Denote $f(\underline{\theta}, \underline{\varepsilon}', \underline{a}, \underline{b})$ and $g_{ij}(\underline{\theta}, \underline{\varepsilon}', \underline{a}, \underline{b}, \underline{c})$ by $f$ and $g_{ij}$ for short. Similar notation is used for the numerator and denominator of the respective rational polynomials.

*Theorem 5:* For an acyclic three-source three-destination unicast network with delays, the input symbols $X_1'^{n+1}$, $X_2'^{n}$, and $X_3'^{n}$ can be exactly recovered at the sinks-1, 2, and 3 from the output symbols $Y_1^{2n+1}$, $Y_2^{2n+1}$, and $Y_3^{2n+1}$ respectively, if the ideal generated by the polynomials $g_{ij}^{(nr)}$ ($i \in \{1,2,..,2n+1\}$ and $j \in \{1,2,..,n\}$), and $\left(1 - \delta f^{(nr)} f^{(dr)} \prod_{(i,j)} g_{ij}^{(dr)}\right)$ does not include 1, where $\delta$ is a variable that takes value from $\overline{\mathbb{F}_p}$.

*Proof:* Proof is given in Appendix I ∎

When the conditions of the above Theorem are satisfied, we say that network alignment is feasible. When network alignment is feasible, throughputs of $\frac{(n+1)}{(2n+1)}$, $\frac{n}{(2n+1)}$, and $\frac{n}{(2n+1)}$ are achieved for the source-destination pairs $S_1 - D_1$, $S_2 - D_2$, and $S_3 - D_3$ respectively. Hence, as $n \to \infty$, a throughput of $1/2$ is achieved for every source-destination pair.

*Remark 6:* If $f(\underline{\theta}, \underline{\varepsilon}', \underline{a}, \underline{b})$ has to be a non-zero polynomial firstly, $V_1$ has to be a full-rank matrix. This is true from the choice of $V_1$. Also, $M_{11}^{-1}M_{21}V_2$, $M_{12}^{-1}M_{22}V_2$ and $M_{13}^{-1}M_{33}V_3$ should also be full-rank. Since $M_{ij}$s are invertible, it is equivalent to checking if $V_2$ and $V_3$ are full-rank. This is also true because $V_1$ is a full-rank matrix and by choosing $A$ and $B$ as matrices that select the first $n$ columns of $V_1$ and the last $n$ columns of $V_1$ respectively, $V_2$ and $V_3$ become full-rank. Hence, the determinants of all the $n \times n$ sub-matrices of $V_2$ and $V_3$ are non-zero polynomials. So, we have atleast ensured that by proper choice of $V_1$, $V_2$ and $V_3$, they are full-rank matrices.

*Remark 7:* Note that the network alignment matrices in Section IV-A can be derived as a special case of the network alignment matrices in Section IV-B. Hence, Theorem 4 can be derived as a special case of Theorem 5. This is explained below. Choose $\underline{\varepsilon}^{(-d_{max})} = \underline{\varepsilon}^{(-d_{max}+1)} = ... = \underline{\varepsilon}^{(2n)} = \underline{\varepsilon}$. Also, choose the variables $\theta_{ij}$ such that $V_1$ in (26) takes the form of $V_1$ in (19). Choose $A$ and $B$, respectively, to be selection matrices which select the first $n$ columns and last $n$ columns of the matrices pre-multiplying them. Let $C = I_n$. Now, it is easy to see that $T_1 V_1 A - V_1 B$ is equal to $Q_1(TV_1 A - V_1 B) = 0$. Now, it can also be easily seen that the full-rank conditions in Theorem 4 are the same as saying that the ideal generated by $(1 - \delta h(\underline{\theta}, \underline{\varepsilon}', \underline{a}, \underline{b}, \underline{c}) f(\underline{\theta}, \underline{\varepsilon}', \underline{a}, \underline{b}))$ should not include 1.

**Case 2:** The min-cut between source-$i$ and sink-$j$ is equal to 0, for some $i \neq j$.

In this case, we have totally 63 possibilities in which one of them is a zero-interference possibility (i.e. min-cut between $S_i$-$D_j$ is equal to zero for all $i \neq j$). Clearly, we need not consider the zero-interference possibility.

We broadly classify the different possibilities into four categories as given in Table II. All the other possibilities involve either permutations of the sources or require minor modifications in the network alignment procedure for one of these categories. For all these categories, network alignment can be done with time-varying LEKs too. But, the only difference with respect to network alignment with time-invariant LEKs would be that the network transfer matrices cannot be diagonalized.

We shall present network alignment for the categories given in Table II with time-invariant LEKs only. We assume the same set-up as in Section IV-A. We shall also assume that $p \nmid 2n+1$ for the same reason as that in Theorem 4.

*Category 1 (Min-Cut between $S_2$-$D_1$ is equal to 0):* This implies that $\hat{M}_{21} = 0$. Let the elements of $V_1$ be given by

$$[V_1]_{ij} = \theta_{ij}, \ i \in \{1, 2, .., 2n+1\}, \ j \in \{1, 2, .., n+1\}, \quad (28)$$

where, $\theta_{ij}$ is a variable that takes values from $\mathbb{F}_q$. Also, let

$$V_2 = \hat{M}_{23}^{-1} \hat{M}_{13} V_1 A \text{ and } V_3 = \hat{M}_{32}^{-1} \hat{M}_{12} V_1 B, \quad (29)$$

where, the elements of the matrices $A$ and $B$, of size $(n+1) \times n$, are given by $[A]_{ij} = a_{ij}$ and $[B]_{ij} = b_{ij}$ respectively ($a_{ij}$ and $b_{ij}$ are variables that take values from $\mathbb{F}_q$). The following theorem provides the conditions under which network alignment can be achieved.

*Theorem 6:* For an acyclic three-source three-destination unicast network with delays, when the min-cut between $S_2$-$D_1$ is equal to 0 and the min-cut between the other sources and destinations are not zero, the input symbols $X_1'^{n+1}$, $X_2'^{n}$, and $X_3'^{n}$ can be exactly recovered at the sinks-1, 2, and 3 from the output symbols $Y_1^{2n+1}$, $Y_2^{2n+1}$, and $Y_3^{2n+1}$ respectively, if

$$\text{Rank}[V_1 \ \hat{M}_{11}^{-1} \hat{M}_{31} V_3] = \text{Rank}[\hat{M}_{12}^{-1} \hat{M}_{22} V_2 \ V_1] = 2n+1,$$
$$\text{Rank}[\hat{M}_{13}^{-1} \hat{M}_{33} V_3 \ V_1] = 2n+1.$$

*Proof:* Proof is given in Appendix J. ∎

When the conditions of the above Theorem are satisfied, throughputs of $\frac{(n+1)}{(2n+1)}$, $\frac{n}{(2n+1)}$, and $\frac{n}{(2n+1)}$ are achieved for the source-destination pairs $S_1 - D_1$, $S_2 - D_2$, and $S_3 - D_3$ respectively. Hence, as $n \to \infty$, a throughput of $1/2$ is achieved for every source-destination pair.

*Category 2 (Min-Cut between $S_2$-$D_1$, $S_3$-$D_1$ and $S_1$-$D_2$ are equal to 0):* This implies that $\hat{M}_{21} = 0$, $\hat{M}_{31} = 0$ and $\hat{M}_{12} = 0$. Let the choice of $V_1$ and $V_2$ be the same as in (28) and (29) respectively, and choose the elements of $V_3$ as

$$[V_3]_{ij} = \delta_{ij}, \ i \in \{1, 2, .., 2n+1\}, \ j \in \{1, 2, .., n\}), \quad (30)$$

where $\delta_{ij}$ is a variable that takes values from $\mathbb{F}_q$. The following theorem provides the conditions under which network alignment can be achieved.

*Theorem 7:* For an acyclic three-source three-destination unicast network with delays, when the min-cut between $S_2$-$D_1$, $S_3$-$D_1$ and $S_1$-$D_2$ are equal to 0 and the min-cut between the other sources and destinations are not zero, the input symbols $X_1'^{(n+1)}$, $X_2'^{n}$, and $X_3'^{n}$ can be exactly recovered at the sinks-1, 2, and 3 from the output symbols $Y_1^{2n+1}$, $Y_2^{2n+1}$, and $Y_3^{2n+1}$ respectively, if

$$\text{Rank}[\hat{M}_{32}^{-1} \hat{M}_{22} V_2 \ V_3] = 2n, \ \text{Rank}[\hat{M}_{33}^{-1} \hat{M}_{13} V_1 \ V_3] = 2n+1.$$

*Proof:* Proof is given in Appendix K. ∎

When the conditions of the above Theorem is satisfied, throughputs of $\frac{(n+1)}{(2n+1)}$, $\frac{n}{(2n+1)}$, and $\frac{n}{(2n+1)}$ are achieved for the source-destination pairs $S_1 - D_1$, $S_2 - D_2$, and $S_3 - D_3$ respectively. Hence, as $n \to \infty$, a throughput of $1/2$ is achieved for every source-destination pair.

*Category 3 (Min-Cut between $S_3$-$D_1$, $S_1$-$D_2$ and $S_2$-$D_3$ are equal to 0):* This implies that $\hat{M}_{31} = 0$, $\hat{M}_{12} = 0$ and $\hat{M}_{23} = 0$. Let the choice of $V_1$ be the same as in (28) and define the elements of $V_2$ and $V_3$ as

$$[V_2]_{ij} = \gamma_{ij}, \ [V_3]_{ij} = \delta_{ij}, \ i \in \{1, 2.., 2n+1\}, j \in \{1, 2.., n\} \quad (31)$$

where, $\gamma_{ij}$ and $\delta_{ij}$ are variables that take values from $\mathbb{F}_q$.

*Theorem 8:* For an acyclic three-source three-destination unicast network with delays, when the min-cut between $S_3$-$D_1$, $S_1$-$D_2$ and $S_2$-$D_3$ are equal to 0 and the min-cut between the other sources and destinations are not zero, the input symbols $X_1'^{n+1}$, $X_2'^{n}$, and $X_3'^{n}$ can be exactly recovered at the sinks-1, 2, and 3 from the output symbols $Y_1^{2n+1}$, $Y_2^{2n+1}$, and $Y_3^{2n+1}$ respectively, if

$$\text{Rank}[V_1 \ \hat{M}_{11}^{-1} \hat{M}_{21} V_2] = \text{Rank}[\hat{M}_{33}^{-1} \hat{M}_{13} V_1 \ V_3] = 2n+1,$$
$$\text{Rank}[\hat{M}_{32}^{-1} \hat{M}_{22} V_2 \ V_3] = 2n.$$

*Proof:* Proof is given in Appendix L. ∎

When the conditions of the above Theorem is satisfied, throughputs of $\frac{(n+1)}{(2n+1)}$, $\frac{n}{(2n+1)}$, and $\frac{n}{(2n+1)}$ are achieved for the source-destination pairs $S_1 - D_1$, $S_2 - D_2$, and $S_3 - D_3$ respectively. Hence, as $n \to \infty$, a throughput of $1/2$ is achieved for every source-destination pair.

*Category 4 (Min-Cut between $S_3$-$D_1$, $S_3$-$D_2$, $S_1$-$D_3$ and $S_2$-$D_3$ are equal to 0):* This implies that $\hat{M}_{31} = 0$, $\hat{M}_{32} = 0$, $\hat{M}_{13} = 0$ and $\hat{M}_{23} = 0$. Here, we can achieve a sum-throughput of 2. Since, $D_3$ is not facing any interference, we can take independent input symbols of source-3, i.e., $X_3'^{2n+1}$, to be of size $(2n+1) \times (2n+1)$ and $V_3$ to be an identity matrix of size $(2n+1) \times (2n+1)$. The independent symbols, $X'^{n+1}_1$ and $X'^{n}_2$ are column vectors of sizes $(n+1) \times 1$ and $n \times 1$ respectively. Let the choice of $V_1$ and $V_2$ be the same as in (28) and (31) respectively. The following theorem provides the conditions under which network alignment can be achieved.

TABLE II
VARIOUS POSSIBILITIES OF MIN-CUTS BETWEEN SOURCE-$i$ AND DESTINATION-$j$ $((i,j) \in \{1,2,3\}|i \neq j)$

| Category No. | Min-Cut between Source-$i$ and Destination-$j$ | | | | | |
|---|---|---|---|---|---|---|
| | $S_2 - D_1$ | $S_3 - D_1$ | $S_1 - D_2$ | $S_3 - D_2$ | $S_1 - D_3$ | $S_2 - D_3$ |
| 1. | 0 | $\geq 1$ | $\geq 1$ | $\geq 1$ | $\geq 1$ | $\geq 1$ |
| 2. | 0 | 0 | 0 | $\geq 1$ | $\geq 1$ | $\geq 1$ |
| 3. | $\geq 1$ | 0 | 0 | $\geq 1$ | $\geq 1$ | 0 |
| 4. | $\geq 1$ | 0 | $\geq 1$ | 0 | 0 | 0 |

*Theorem 9:* For an acyclic three-source three-destination unicast network with delays, when the min-cut between $S_3$-$D_1$, $S_3$-$D_2$, $S_1$-$D_3$, and $S_2$-$D_3$ are equal to 0 and the min-cut between the other sources and destinations are not zero, the input symbols $X'^{n+1}_1$, $X'^{n}_2$, and $X'^{2n+1}_3$ can be exactly recovered at the sinks-1, 2 and 3 from the output symbols $Y^{2n+1}_1$, $Y^{2n+1}_2$ and $Y^{2n+1}_3$ respectively, if

$$\text{Rank}[V_1 \quad \hat{M}^{-1}_{11}\hat{M}_{21}V_2] = \text{Rank}[\hat{M}^{-1}_{12}\hat{M}_{22}V_2 \quad V_1] = 2n+1.$$

*Proof:* Proof is given in Appendix M. ∎

When the conditions of the above Theorem is satisfied, throughputs of $\frac{(n+1)}{(2n+1)}$, $\frac{n}{(2n+1)}$, and $\frac{(2n+1)}{(2n+1)}$ are achieved for the source-destination pairs $S_1 - D_1$, $S_2 - D_2$, and $S_3 - D_3$ respectively. Hence, as $n \to \infty$, a throughput of $1/2$ is achieved for $S_1 - D_1$, $S_2 - D_2$ and a throughput of 1 is easily achieved for $S_3 - D_3$.

*Remark 8:* In all the above four categories, the choices of $V_1$, $V_2$ and $V_3$ were such that we could atleast ensure that $V_1$, $V_2$ and $V_3$ were full-rank, which were necessary to satisfy the network-alignment conditions.

*Remark 9:* In Category 4, a sum-throughput of close to 2 is achieved as $n \to \infty$. For acyclic networks without delays, this can be easily achieved by time-sharing the network equally between source-1 and source-2. But it is not clear how such a sum-throughput can be achieved for arbitrary acyclic networks with delays whereas, our method provides a scheme that can achieve it.

## V. DISCUSSION

Though the transform method was originally claimed to be applicable for ayclic networks having $M(D)$ whose elements are only polynomial functions in $D$, it can also be applied to networks having $M(D)$ whose elements are rational functions in $D$ by multiplying by the LCM of all the denominators of the rational functions, at all the sinks. This gives a finite $d_{max}$. The same applies to cylic networks too.

Network alignment for the three source-three destination unicast network with delays, discussed in this paper, can be extended to the case where each source-destination pair has a min-cut greater than one. We are currently working on it. An interesting dierction of future research is extending the network alignment to the case of arbitary number of sources and destinations with arbitrary message demands.

ACKNOWLEDGEMENT

This work was supported partly by the DRDO-IISc program on Advanced Research in Mathematical Engineering through a research grant as well as the INAE Chair Professorship grant to B. S. Rajan.

## APPENDIX A
## PROOF OF THEOREM 1

*Proof:*

$$A \begin{bmatrix} I_\mu \\ \alpha^j I_\mu \\ \alpha^{2j} I_\mu \\ \vdots \\ \alpha^{(n-1)j} I_\mu \end{bmatrix}_{n\mu \times \mu} = \begin{bmatrix} \sum_{i=0}^L \alpha^{ij} A_i \\ \sum_{i=0}^L \alpha^{(i+1)j} A_i \\ \vdots \\ \sum_{i=0}^L \alpha^{(i+n-1)j} A_i \end{bmatrix}_{n\nu \times \mu}$$

$$= \begin{bmatrix} I_\nu \\ \alpha^j I_\nu \\ \alpha^{2j} I_\nu \\ \vdots \\ \alpha^{(n-1)j} I_\nu \end{bmatrix}_{n\nu \times \nu} \left( \sum_{i=0}^L \alpha^{ij} A_i \right)$$
(32)

The inverse of the matrix $F$ is given by

$$F^{-1} = n^{-1} \begin{bmatrix} 1 & 1 & 1 & \cdots & 1 \\ 1 & \alpha^{-1} & \alpha^{-2} & \cdots & \alpha^{-(n-1)} \\ 1 & \alpha^{-2} & \alpha^{-4} & \cdots & \alpha^{-2(n-1)} \\ \vdots & \ddots & \ddots & \ddots & \ddots \\ 1 & \alpha^{-(n-1)} & \alpha^{-2(n-1)} & \cdots & \alpha^{-(n-1)(n-1)} \end{bmatrix}.$$

Note that $F^{-1}$ exists [9]. Now, $Q_\mu$ can also be written as $Q_\mu = F \otimes I_\mu$ (*i.e.* Kronecker product of $F$ and $I_\mu$). Similarly $Q_\nu = F \otimes I_\nu$. From (32), we have

$$AQ_\mu = Q_\nu \hat{A}.$$

Now, $det(Q_\mu) = [det(F)]^\mu [det(I_\mu)]^n \neq 0$ and $Q_\mu^{-1} = F^{-1} \otimes I_\mu$ ($\because Q_\mu Q_\mu^{-1} = (F \otimes I_\mu)(F^{-1} \otimes I_\mu) = (FF^{-1}) \otimes I_\mu = I_{n\mu}$). So,

$$A = Q_\nu \hat{A} Q_\mu^{-1}.$$

Hence the theorem is proved. ∎

## APPENDIX B
## PROOF OF THEOREM 2

*Proof:* If both the conditions are satisfied after the assignment of values to $\underline{\varepsilon}$, then sink-$j$ can invert $[\hat{M}_{i_1 j}^{(k)}(l_{i_1}) \ \hat{M}_{i_2 j}^{(k)}(l_{i_2}) \ \cdots \ \hat{M}_{i_{s'} j}^{(k)}(l_{i_{s'}})]$ matrix and decode the required input symbols without any interference.

If Condition 1) is not satisfied, then sink receives superposition of required information and interference from other input symbols, which it cannot distinguish.

If Condition 2) is not satisfied, then sink cannot invert the matrix $[\hat{M}_{i_1 j}^{(k)}(l_{i_1}) \ \hat{M}_{i_2 j}^{(k)}(l_{i_2}) \ \cdots \ \hat{M}_{i_{s'} j}^{(k)}(l_{i_{s'}})]$ which is necessary for decoding the input symbols. ∎

## APPENDIX C
## PROOF OF LEMMA 2

*Proof:* Following the terminology developed so far, for some $n >> d_{max}$ and for each $0 \leq t \leq n-1$, let

$$\underline{X}^{(t)} = \begin{bmatrix} \underline{X_1}^{(t)} \\ \underline{X_2}^{(t)} \\ \vdots \\ \underline{X_s}^{(t)} \end{bmatrix}.$$

Then, by (6), (14) and the structure of the $\hat{M}_{ij}^{(t)}$ matrices, we have for $0 \leq t \leq n-1$,

$$\underline{Y_j}^{(t)} = \left( \sum_{d=0}^{d_{max}} \alpha^{d(n-1-t)} M_j^{(d)} \right) \underline{X}^{(t)}, \quad (33)$$

where $M_j^{(d)}$ is a $\nu_j \times \mu$ matrix over $\mathbb{F}_q$ (considered as a subfield of $\mathbb{F}_{q^a}$ such that

$$M_j(D) = \sum_{d=0}^{d_{max}} M_j^{(d)} D^d. \quad (34)$$

We define a collection of ring homomorphisms $\phi_t : \mathbb{F}_q(D) \to \mathbb{F}_{q^a}$ for $0 \leq t \leq n-1$, given by $\phi_t(D) = \alpha^t$. For some matrix $P(D)$ over $\mathbb{F}_q(D)$, we also define $\phi_t(P(D))$ to be equal to the matrix $P$ with elements in $\mathbb{F}_{q^a}$ that are the $\phi_t$-images of the corresponding elements of $P(D)$. Then, from (33) and (34), we have

$$\underline{Y_j}^{(n-1-t)} = \phi_t(M_j(D)) \underline{X}^{(n-1-t)}, \quad (35)$$

for $0 \leq t \leq n-1$. Clearly, the zero-interference conditions satisfied in the $M_j(D)$ matrices continue to hold in the $\phi_t(M_j(D))$ matrices, for any $0 \leq t \leq n-1$ and for any sink-$j$. Having satisfied the zero-interference conditions, to recover the source processes demanded by each sink-$j$ at time instant $n-1-t$, the invertibility conditions also have to be satisfied, i.e.,

$$\prod_{j=1}^r det\left(\phi_t(M'_j(D))\right) \neq 0, \quad (36)$$

where $M'_j(D)$ is the square submatrix of $M_j(D)$ indicating the source processes that are demanded by sink-$j$. But then, we have

$$det\left(\phi_t(M'_j(D))\right) = \phi_t(det(M'_j(D))) \quad (37)$$

and thus

$$\prod_{j=1}^r det\left(\phi_t(M'_j(D))\right) = \prod_{j=1}^r \phi_t\left(det(M'_j(D))\right)$$
$$= \phi_t\left(\prod_{j=1}^r det(M'_j(D))\right)$$
$$= \phi_t(f(D))$$
$$= f(\alpha^t),$$

where $f(D)$ is as defined in (15). Clearly, $f(\alpha^t) \neq 0$ implies that (36) is satisfied and the source processes demanded at each sink can be recovered at time instant $n-1-t$ in the transform approach. Similarly, if the sink demands are satisfied at time instant $n-1-t$ in the transform approach, clearly we must have $f(\alpha^t) \neq 0$. This holds for $0 \leq t \leq n-1$, thus proving the lemma. ∎

## APPENDIX D
## PROOF OF THEOREM 3

*Proof: If part:*

Let $\mathbb{F}_{p^m}$ be the field over which the feasible network code has been obtained for $(\mathcal{G},\mathcal{C})$. Consider the polynomial $f(D)$ (given by (15)) with coefficients from $\mathbb{F}_{p^m}$. Let $\mathbb{F}_{p^{m'}}$ be the splitting field of this polynomial, i.e., a suitable smallest extension field of $\mathbb{F}_{p^m}$ in which $f(D)$ splits into linear factors.

Let
$$p^{m'} - 1 = \prod_{b=1}^{b=k} p_b^{m'_b},$$

where each $p_b$ is some prime and $m_b$ is some positive integer.

By Lemma 2, the choice of $\alpha$ to be used for the DFT operations should be such that $f(\alpha^t) \neq 0$, for any $0 \leq t \leq n-1$. We now show that such an $\alpha$ exists and can be chosen.

Let $\mathbb{F}_{p^{m''}}$ be an extension field of $\mathbb{F}_{p^{m'}}$. Clearly, $\left(p^{m'}-1\right) \mid \left(p^{m''}-1\right)$. However, we further demand that $\mathbb{F}_{p^{m''}}$ is such that

$$p^{m''} - 1 = \prod_{b=1}^{b=k} p_b^{m''_b} \prod_{c=1}^{c=k'} p_c^{m''_c}, \quad (38)$$

where each $p_c$ is some prime and $m''_b$ and $m''_c$ are some positive integer such that $p_b \neq p_c$ for $1 \leq b \leq k$ and $1 \leq c \leq k'$. Note that $m''_b \geq m_b$ for $1 \leq b \leq k$. Such extensions of $\mathbb{F}_{p^{m''}}$ can indeed be obtained. For example, $\mathbb{F}_{p^{m''}}$ can be considered to be the smallest field which contains $\mathbb{F}_{p^{m'}}$ and $\mathbb{F}_{p^{\tilde{m}}}$, $\tilde{m}$ being some positive integer coprime with $m'$. Then clearly $\mathbb{F}_{p^{m''}}$ is such that (38) holds.

Following the notations of Section III, we now pick $\alpha \in \mathbb{F}_{p^{m''}}$ (where $m''$ satisfies (38)) such that the following condition holds

- The cyclic subgroup $\{1, \alpha, ..., \alpha^{n-1}\}$ of $\mathbb{F}_{p^{m''}} \setminus \{0\}$ with order $n(n > 1)$ is such that $n$ and $\prod_{b=1}^{b=k} p_b^{m''_b}$ are coprime.

Such an $\alpha$ can be obtained by choosing $\alpha$ from the subgroup of $\mathbb{F}_{p^{m''}} \setminus \{0\}$ with $n = \prod_{c=1}^{c=k'} p_c^{m''_c}$ elements. We now claim that using such an $\alpha$ for the DFT will result in a feasible transform network code for $(\mathcal{G},\mathcal{C})$. The proof is as follows.

We first note that the zero-interference conditions are satisfied irrespective of the choice of $\alpha$ in the DFT operations. As for the invertibility conditions, by Lemma 2, it is clear that as long as $f(\alpha^t) \neq 0$ for $0 \leq t \leq n-1$, we have a feasible transform network code for $(\mathcal{G},\mathcal{C})$. Suppose $f(\alpha^t) = 0$ for some $1 \leq t \leq n-1$. Let $n_t$ be the order of $\alpha^t$, i.e. the number of elements in the cyclic group generated by $\alpha^t$. Then $n_t | n$ and also $n_t | \prod_{b=1}^{b=k} p_b^{m''_b}$ as $\alpha^t \in \mathbb{F}_{p^{m'}}$ is a zero of $f(D)$. However this leads to a contradiction as $n$ shares no common prime factor with $\prod_{b=1}^{b=k} p_b^{m''_b}$. Thus no $\alpha^t$, $1 \leq t \leq n-1$, can be a zero of $f(D)$. This, coupled with the given fact that $f(1) \neq 0$, proves the claim and hence the if part of the theorem.

*Only If part:*

Let $\mathbb{F}_q$ be the field over which a feasible transform network code has been defined for $(\mathcal{G},\mathcal{C})$, i.e., there exists a choice of LEKs and for $\alpha$ from $\mathbb{F}_q$ using which the zero-interference and the invertibility constraints have been satisfied in the transform domain. Note that a choice for the LEKs implies that the matrices $M_j(D)$ given by (6) are well defined. We will now prove that the invertibility and the zero-interference constraints also hold in these $M_j(D)$ matrices for all sinks, i.e. for $1 \leq j \leq r$.

We first prove the invertibility conditions. Towards that end, let $\hat{M}_j^{(n-1)}$ be defined as the $\nu_j \times \mu$ transfer matrix at time instant $n-1$ from all the sources to sink-$j$ in the transform approach, i.e.,

$$\hat{M}_j^{(n-1)} = \left[\hat{M}_{1j}^{(n-1)} \hat{M}_{2j}^{(n-1)}...\hat{M}_{sj}^{(n-1)}\right]. \quad (39)$$

By the structure of the $\hat{M}_{ij}^{(n-1)}$ matrices, we have $\hat{M}_j^{(n-1)} = \sum_{d=0}^{d=d_{max}} M_j^{(d)} = M_j(D)|_{D=1}$. Let $\hat{M}_j^{'(n-1)}$ be the submatrix of $\hat{M}_j^{(n-1)}$ which is known to be invertible, as it is given that the invertibility conditions for the transform network code are all satisfied.

The invertibility conditions for sink-$j$ of the usual (non-transform) network code for $(\mathcal{G},\mathcal{C})$ demand a suitable submatrix $M'_j(D)$ of the matrix $M_j(D)$ to be invertible. Note however that $M'_j(D)|_{D=1} = \hat{M}_j^{'(n-1)}$, by (39). Therefore, we have $det\left(\hat{M}_j^{'(n-1)}\right) = det\left(M'_j(D)|_{D=1}\right) \neq 0$. As in (37), we have $det\left(M'_j(D)\right)|_{D=1} = det\left(M'_j(D)|_{D=1}\right) \neq 0$. Therefore, $det\left(M'_j(D)\right) \neq 0$, i.e., $det\left(M'_j(D)\right)$ is a non-zero polynomial in $D$. Because the choice of the sink was arbitrary, it is clear that the invertibility conditions hold for each sink in the usual network code for $(\mathcal{G},\mathcal{C})$. By (15), we also have $(D-1) \nmid f(D)$.

We now prove the zero-interference conditions. The zero-interference conditions in the transform domain can be interpreted as follows. Having ordered the input processes at the source-$i$, suppose the sink-$j$ does not demand the $k^{th}$ process from the source-$i$. Then the matrix $\hat{M}_{ij}$ is such that $k^{th}$ column of $\hat{M}_{ij}^{(t)}$ is an all-zero column for all $0 \leq t \leq n-1$. To prove that the zero-interference conditions continue to hold in the usual network code for $(\mathcal{G},\mathcal{C})$, we must then prove that for each source-$i$, each particular sink-$j$ and each $k$ (such that the $k^{th}$ input process at source-$i$ is not demanded at sink-$j$, the $k^{th}$ columns of $M_{ij}^{(d)}$ matrices are all-zero for $0 \leq d \leq d_{max}$ where $M_{ij}^{(d)}, 0 \leq d \leq d_{max}$ are matrices such that

$$M_{ij}(D) = \sum_{d=0}^{d_{max}} M_{ij}^{(d)} D^d.$$

This is seen by observing the structure of the $M_{ij}$ matrix, which is defined by (11). Using Theorem 1 and with $\beta_a = \alpha^a$, we have (40) (shown at the top of the next page). Comparing

$$M_{ij} = Q_{\nu_j} \hat{M}_{ij} Q_{\mu_i}^{-1}$$

$$= \begin{bmatrix} I_{\nu_j} & I_{\nu_j} & I_{\nu_j} & \cdots & I_{\nu_j} \\ I_{\nu_j} & \beta_1 I_{\nu_j} & \beta_1^2 I_{\nu_j} & \cdots & \beta_1^{n-1} I_{\nu_j} \\ \vdots & \vdots & \vdots & \cdots & \vdots \\ I_{\nu_j} & \beta_{n-1} I_{\nu_j} & \beta_{n-1}^2 I_{\nu_j} & \cdots & \beta_{n-1}^{n-1} I_{\nu_j} \end{bmatrix} \begin{bmatrix} \hat{M}_{ij}^{(n-1)} & 0 & 0 & \cdots & 0 \\ 0 & \hat{M}_{ij}^{(n-2)} & 0 & \cdots & 0 \\ \vdots & \vdots & \vdots & \cdots & \vdots \\ 0 & 0 & 0 & \cdots & \hat{M}_{ij}^{(0)} \end{bmatrix} \begin{bmatrix} I_{\mu_i} & I_{\mu_i} & I_{\mu_i} & \cdots & I_{\mu_i} \\ I_{\mu_i} & \beta_1^{-1} I_{\mu_i} & \beta_1^{-2} I_{\mu_i} & \cdots & \beta_1^{-(n-1)} I_{\mu_i} \\ \vdots & \vdots & \vdots & \cdots & \vdots \\ I_{\mu_i} & \beta_{n-1}^{-1} I_{\mu_i} & \beta_{n-1}^{-2} I_{\mu_i} & \cdots & \beta_{n-1}^{-(n-1)} I_{\mu_i} \end{bmatrix}$$

$$= \begin{bmatrix} \sum_{t=0}^{n-1} \hat{M}_{ij}^{(t)} & \sum_{t=0}^{n-1} \beta_{n-1-t}^{-1} \hat{M}_{ij}^{(t)} & \cdots & \sum_{t=0}^{n-1} \beta_{n-1-t}^{-(n-1)} \hat{M}_{ij}^{(t)} \\ \sum_{t=0}^{n-1} \beta_1^{n-1-t} \hat{M}_{ij}^{(t)} & \sum_{t=0}^{n-1} \hat{M}_{ij}^{(t)} & \cdots & \sum_{t=0}^{n-1} \beta_1^{n-1-t} \beta_1^{-(n-1)} \hat{M}_{ij}^{(t)} \\ \vdots & \vdots & \cdots & \vdots \\ \sum_{t=0}^{n-1} \beta_{n-1}^{n-1-t} \hat{M}_{ij}^{(t)} & \sum_{t=0}^{n-1} \beta_{n-1}^{n-1-t} \beta_{n-1-t}^{-1} \hat{M}_{ij}^{(t)} & \cdots & \sum_{t=0}^{n-1} \hat{M}_{ij}^{(t)} \end{bmatrix}. \quad (40)$$

---

the submatrices of $M_{ij}$ from (11) and (40), we see that if the $k^{th}$ column of the $\hat{M}_{ij}^{(t)}$ matrices is all-zero for all $0 \leq t \leq n-1$, then the $k^{th}$ columns of $M_{ij}^{(d)}$ matrices are all-zero for $0 \leq d \leq d_{max}$. As the choice of source-$i$ and sink-$j$ are arbitrary, it is clear that the zero-interference conditions continue to hold in the $M_{ij}(D)$ matrices for all $1 \leq i \leq s$ and $1 \leq j \leq r$. This proves the only if part of the theorem and hence the theorem is proved. ∎

## APPENDIX E
## PROOF OF LEMMA 3

*Proof:* Consider $M_{ij}$ as defined in (11) which is a circulant matrix of size $(2n+1) \times (2n+1)$. Note that the diagonal elements of $\hat{M}_{ij}$, i.e., $\hat{M}_{ij}^{(k)}$ ($k \in 0, 1, .., 2n$), are the eigen values of the matrix $M_{ij}$. Also, note that the eigen values are equal to $(2n+1)$-point finite-field DFT of the first row of $M_{ij}$. Since, the min-cut from source-$i$ to sink-$j$ is equal to 1, by Menger's Theorem, there exists exactly one link-disjoint directed path from source-$i$ to sink-$j$. Let such a directed path consist of links $e_1, e_2, .., e_t$. Now, we can assign the values $\alpha_{1,e_1} = 1$, $\beta_{e_i,e_{i+1}} = 1$ ($i \in \{1,2,..,t-1\}$), $\epsilon_{e_t,1} = 1$ and assign values of 0 to all the other LEKs. By, such an assignment of values to the LEKs, exactly one among $M_{ij}^{(0)}$, $M_{ij}^{(1)}, .., M_{ij}^{(d_{max})}$ is equal to 1. This implies that all the eigen values of $M_{ij}$ are non-zero. Hence, the diagonal elements of $\hat{M}_{ij}$ are non-zero polynomials in $\underline{\varepsilon}$ and so is its determinant. ∎

## APPENDIX F
## PROOF OF LEMMA 4

*Proof: If part:* Euler's theorem [14] states that if two positive integers $a$ and $b$ are coprime then, $b$ divides $a^{\phi(b)} - 1$ where $\phi$ represents the Euler's totient function. If $2n+1 < p$ then, $2n+1$ and $p$ are coprime. If $2n+1 \geq p$ then, $p$ and $2n+1$ are coprime iff $p$ does not divide $2n+1$. Hence, by Euler's theorem, $2n+1 | p^{\phi(2n+1)} - 1$ if $p \nmid 2n+1$. Thus if $p \nmid 2n+1$ then, $2n+1 | p^m - 1$, for all $m$ such that $\phi(2n+1) | m$.
*Only If part:* If $2n+1$ divides $p^m - 1$ for some positive integer $m$ then, $p^m - 1 = r(2n+1)$ for some positive integer $r$. So, $p^m - (2n+1)r = 1$ which means that $p$ and $2n+1$ must be coprime. Since $p$ is prime, $p \nmid 2n+1$. ∎

## APPENDIX G
## PROOF OF THEOREM 4

*Proof:* To exactly recover $X_1'^{n+1}$, $X_2'^{n}$ and $X_3'^{n}$ at the sinks-1, 2 and 3 respectively, it is sufficient that the following network alignment conditions are satisfied.

$$\hat{M}_{21} V_2 = \hat{M}_{31} V_3 \quad (41)$$
$$\hat{M}_{32} V_3 \subset \hat{M}_{12} V_1 \quad (42)$$
$$\hat{M}_{23} V_2 \subset \hat{M}_{13} V_1 \quad (43)$$
$$\text{Rank}[\hat{M}_{11} V_1 \quad \hat{M}_{21} V_2] = 2n+1$$
$$\Leftrightarrow \text{Rank}[V_1 \quad \hat{M}_{11}^{-1} \hat{M}_{21} V_2] = 2n+1 \quad (44)$$
$$\text{Rank}[\hat{M}_{22} V_2 \quad \hat{M}_{12} V_1] = 2n+1$$
$$\Leftrightarrow \text{Rank}[\hat{M}_{12}^{-1} \hat{M}_{22} V_2 \quad V_1] = 2n+1 \quad (45)$$
$$\text{Rank}[\hat{M}_{33} V_3 \quad \hat{M}_{13} V_1] = 2n+1$$
$$\Leftrightarrow \text{Rank}[\hat{M}_{13}^{-1} \hat{M}_{33} V_3 \quad V_1] = 2n+1 \quad (46)$$

Note that from Lemma 3, inverse of $\hat{M}_{ij} \ \forall \ (i,j) \in \{1,2,3\}$ is well-defined. It is easily seen that the choice of $V_1$, $V_2$, and $V_3$ in (19)-(21) satisfy the conditions (41)-(43). Suppose that (44)-(46) are satisfied. Let

$$f_1(\underline{\varepsilon}) = det([V_1 \quad \hat{M}_{11}^{-1} \hat{M}_{21} V_2])$$
$$f_2(\underline{\varepsilon}) = det([\hat{M}_{12}^{-1} \hat{M}_{22} V_2 \quad V_1])$$
$$f_3(\underline{\varepsilon}) = det([\hat{M}_{13}^{-1} \hat{M}_{33} V_3 \quad V_1])$$
$$f_4(\underline{\varepsilon}) = \prod_{(i,j) \in \{1,2,3\}} det(M_{ij})$$
$$f(\underline{\varepsilon}) = \prod_{i=1}^{4} f_i(\underline{\varepsilon}).$$

Since $f_1(\underline{\varepsilon})$, $f_2(\underline{\varepsilon})$ and $f_3(\underline{\varepsilon})$ are non-zero polynomials in $\underline{\varepsilon}$, $f(\underline{\varepsilon})$ is also a non-zero polynomial in $\underline{\varepsilon}$. Hence, by Lemma 1 in [3], for a sufficiently large field size, there exists an assignment of values to $\underline{\varepsilon}$ such that the network alignment conditions are satisfied. Since $p \nmid 2n+1$, by Lemma 4, for a sufficiently large $m$ (in particular, $m$ such that $\phi(2n+1) | m$ where $\phi$ represents the Euler's totient function), there exists an assignment of values to $\underline{\varepsilon}$ such that the network alignment conditions are satisfied. Hence, the theorem is proved. ∎

## APPENDIX H
## PROOF OF LEMMA 5

*Proof:* If we assign $\underline{\varepsilon}^{(-d_{max})} = \underline{\varepsilon}^{(-d_{max}+1)} = \ldots = \underline{\varepsilon}^{(2n)} = \underline{\varepsilon}$, $M_{ij}$ in (25) becomes a circulant matrix and hence can be diagonalized as shown in Theorem 1. Further, in lemma 3 we proved that the determinant of the diagonalized matrix is a non-zero polynomial in $\underline{\varepsilon}$. So, the determinant of the circulant matrix is also a non-zero polynomial in $\underline{\varepsilon}$. Hence, the determinant of $M_{ij}$ is a non-zero polynomial in $\underline{\varepsilon}'$. ∎

## APPENDIX I
## PROOF OF THEOREM 5

*Proof:* To exactly recover $X_1'^{n+1}$, $X_2'^n$ and $X_3'^n$ at the sinks-1, 2 and 3 respectively, it is sufficient that the following network alignment conditions are satisfied.

$$\text{Span}(M_{21}V_2) = \text{Span}(M_{31}V_3) \tag{47}$$
$$\text{Span}(M_{32}V_3) \subset \text{Span}(M_{12}V_1) \tag{48}$$
$$\text{Span}(M_{23}V_2) \subset \text{Span}(M_{13}V_1) \tag{49}$$
$$\text{Rank}[M_{11}V_1 \quad M_{21}V_2] = 2n+1$$
$$\Leftrightarrow \text{Rank}[V_1 \quad M_{11}^{-1}M_{21}V_2] = 2n+1 \tag{50}$$
$$\text{Rank}[M_{22}V_2 \quad M_{12}V_1] = 2n+1$$
$$\Leftrightarrow \text{Rank}[M_{12}^{-1}M_{22}V_2 \quad V_1] = 2n+1 \tag{51}$$
$$\text{Rank}[M_{33}V_3 \quad M_{13}V_1] = 2n+1$$
$$\Leftrightarrow \text{Rank}[M_{13}^{-1}M_{33}V_3 \quad V_1] = 2n+1 \tag{52}$$

The choice of $V_1$, $V_2$ and $V_3$ ensures that the conditions (48) and (49) are satisfied. To satisfy (47), we have to ensure that

$$M_{31}^{-1}M_{21}V_2 = M_{32}^{-1}M_{12}V_1BC$$
$$\Leftrightarrow M_{31}^{-1}M_{21}M_{23}^{-1}M_{13}V_1A = M_{32}^{-1}M_{12}V_1BC$$
$$\Leftrightarrow T_1V_1A = V_1BC \tag{53}$$

is satisfied. In order to satisfy (53), every element of $T_1V_1A$ must be equal to every element of $V_1BC$, i.e.,

$$g_{ij} = 0, \ \forall \ i \in \{1,2,..,2n+1\}, \ j \in \{1,2,..,n\}.$$

Hence, to satisfy (50)-(53) we need to find an assignment to the variables - $\underline{\theta}$, $\underline{\varepsilon}'$, $\underline{a}$ and $\underline{b}$, such that $f \neq 0$ and $g_{ij} = 0$, $\forall \ i \in \{1,2,..,2n+1\}, \ j \in \{1,2,..,n\}$. This means that there must exist an assignment such that $f^{(nr)} \neq 0$ and $g_{ij}^{(nr)} = 0$. After the assignment to the variables, we require that $f^{(dr)} \neq 0$ and $g_{ij}^{(dr)} \neq 0$ as dividing by zero is prohibited. In order to formulate this as an algebraic problem, introduce a new variable $\delta$ and consider the polynomial $\left(1 - \delta f^{(nr)}f^{(dr)}\prod_{(i,j)} g_{ij}^{(dr)}\right)$. From Weak Nullstellensatz [15], an assignment to the variables - $\delta$, $\underline{\theta}$, $\underline{\varepsilon}'$, $\underline{a}$, $\underline{b}$ and $\underline{c}$ exist such that $g_{ij}^{(nr)} = 0$, for all $(i,j)$, and $\left(1 - \delta f^{(nr)}f^{(dr)}\prod_{(i,j)} g_{ij}^{(dr)}\right) = 0$ iff 1 does not belong to the ideal generated by the polynomials $g_{ij}^{(nr)}$ for all $(i,j)$ and $\left(1 - \delta f^{(nr)}f^{(dr)}\prod_{(i,j)} g_{ij}^{(dr)}\right)$. ∎

## APPENDIX J
## PROOF OF THEOREM 6

*Proof:* Let $\underline{\theta} = \{\theta_{ij} \ \forall \ (i,j)\}$, $\underline{a} = \{a_{ij} \ \forall \ (i,j)\}$ and $\underline{b} = \{b_{ij} \ \forall \ (i,j)\}$. To exactly recover $X_1'^{n+1}$, $X_2'^n$ and $X_3'^n$ at the sinks-1, 2 and 3 respectively, it is sufficient that the following network alignment conditions are satisfied.

$$\text{Span}(\hat{M}_{32}V_3) \subset \text{Span}(\hat{M}_{12}V_1) \tag{54}$$
$$\text{Span}(\hat{M}_{23}V_2) \subset \text{Span}(\hat{M}_{13}V_1) \tag{55}$$
$$\text{Rank}[\hat{M}_{11}V_1 \quad \hat{M}_{31}V_3] = 2n+1$$
$$\Leftrightarrow \text{Rank}[V_1 \quad \hat{M}_{11}^{-1}\hat{M}_{31}V_3] = 2n+1 \tag{56}$$
$$\text{Rank}[\hat{M}_{22}V_2 \quad \hat{M}_{12}V_1] = 2n+1$$
$$\Leftrightarrow \text{Rank}[\hat{M}_{12}^{-1}\hat{M}_{22}V_2 \quad V_1] = 2n+1 \tag{57}$$
$$\text{Rank}[\hat{M}_{33}V_3 \quad \hat{M}_{13}V_1] = 2n+1$$
$$\Leftrightarrow \text{Rank}[\hat{M}_{13}^{-1}\hat{M}_{33}V_3 \quad V_1] = 2n+1 \tag{58}$$

It is easily seen that the choice of $V_2$ and $V_3$, in (29), satisfy the conditions (54) and (55). Suppose, (56)-(58) are satisfied. Now, let

$$f_1(\underline{\varepsilon},\underline{\theta},\underline{a}) = det([\hat{M}_{12}^{-1}\hat{M}_{22}V_2 \quad V_1])$$
$$f_2(\underline{\varepsilon},\underline{\theta},\underline{b}) = det([V_1 \quad \hat{M}_{11}^{-1}\hat{M}_{31}V_3])$$
$$f_3(\underline{\varepsilon},\underline{\theta},\underline{b}) = det([\hat{M}_{13}^{-1}\hat{M}_{33}V_3 \quad V_1])$$
$$f_4(\underline{\varepsilon}) = \prod_{(i,j)\in\{1,2,3\}|(i,j)\neq(2,1)} det(M_{ij})$$

Let $f(\underline{\varepsilon},\underline{\theta},\underline{a}),\underline{b}) = f_1(\underline{\varepsilon},\underline{\theta},\underline{a})\prod_{i=2}^{3} f_4(\underline{\varepsilon})f_i(\underline{\varepsilon},\underline{\theta},\underline{b})$. Since, $f_1(\underline{\varepsilon},\underline{\theta},\underline{a})$, $f_2(\underline{\varepsilon},\underline{\theta},\underline{b})$ and $f_3(\underline{\varepsilon},\underline{\theta},\underline{b})$ are non-zero polynomials, $f(\underline{\varepsilon},\underline{\theta},\underline{a},\underline{b})$ is also a non-zero polynomial in $\underline{\varepsilon}$. Hence, by Lemma 1 in [3] and Lemma 4, for a sufficiently large field size, there exists an assignment of values to variables $\underline{\varepsilon}$, such that the network alignment conditions are satisfied. Hence, the theorem is proved. ∎

## APPENDIX K
## PROOF OF THEOREM 7

*Proof:* Let $\underline{\delta} = \{\delta_{ij} \ \forall \ (i,j)\}$. To exactly recover $X_1'^{n+1}$, $X_2'^n$ and $X_3'^n$ at the sinks-1, 2 and 3 respectively, it is sufficient that the following network alignment conditions are satisfied.

$$\text{Span}(\hat{M}_{23}V_2) \subset \text{Span}(\hat{M}_{13}V_1) \tag{59}$$
$$\text{Rank}[\hat{M}_{11}V_1] = n+1 \tag{60}$$
$$\text{Rank}[\hat{M}_{22}V_2 \quad \hat{M}_{32}V_3] = 2n$$
$$\Leftrightarrow \text{Rank}[\hat{M}_{32}^{-1}\hat{M}_{22}V_2 \quad V_3] = 2n \tag{61}$$
$$\text{Rank}[\hat{M}_{13}V_1 \quad \hat{M}_{33}V_3] = 2n+1$$
$$\Leftrightarrow \text{Rank}[\hat{M}_{33}^{-1}\hat{M}_{13}V_1 \quad V_3] = 2n+1 \tag{62}$$

It is easily seen that the choice of $V_2$ as in (29), satisfies the condition in (59). Since $V_1$ is full-rank and $\hat{M}_{11}$ is invertible

(from Lemma 3), (60) is also satisfied. Suppose (61) and (62) are satisfied. Let $f_1(\underline{\varepsilon}, \underline{\theta})$ denote the determinants of all the $n \times n$ sub-matrices of $\hat{M}_{11}V_1$. Also, let $f_2(\underline{\varepsilon}, \underline{\theta}, \underline{\delta}, \underline{a})$ denote the determinants of all the $2n \times 2n$ sub-matrices of $[\hat{M}_{32}^{-1}\hat{M}_{22}V_2 \quad V_3]$. Now, define

$$f_3(\underline{\varepsilon}, \underline{\theta}, \underline{\delta}) = [\hat{M}_{33}^{-1}\hat{M}_{13}V_1 \quad V_3]$$
$$f_4(\underline{\varepsilon}) = \prod_{(i,j)\in\{1,2,3\}|(i,j)\neq\{(2,1),(3,1),(1,2)\}} det(M_{ij}).$$

Let $f(\underline{\varepsilon}, \underline{\theta}, \underline{\delta}, \underline{a}) = f_1(\underline{\varepsilon}, \underline{\theta})f_2(\underline{\varepsilon}, \underline{\theta}, \underline{\delta}, \underline{a})f_3(\underline{\varepsilon}, \underline{\theta}, \underline{\delta})f_4(\underline{\varepsilon})$. Since, $f_1(\underline{\varepsilon}, \underline{\theta})$, $f_2(\underline{\varepsilon}, \underline{\theta}, \underline{\delta}, \underline{a})$ and $f_3(\underline{\varepsilon}, \underline{\theta}, \underline{\delta}, \underline{a})$ are non-zero polynomials, $f(\underline{\varepsilon}, \underline{\theta}, \underline{\delta}, \underline{a})$ is also a non-zero polynomial. Hence, by Lemma 1 in [3] and Lemma 4, for a sufficiently large field size, there exists an assignment of values to variables $\underline{\varepsilon}$, $\underline{\theta}$, $\underline{\delta}$ and $\underline{a}$ such that the network alignment conditions are satisfied. Hence, the theorem is proved. ∎

## APPENDIX L
## PROOF OF THEOREM 8

*Proof:* To exactly recover $X_1'^{n+1}$, $X_2'^{n}$ and $X_3'^{n}$ at the sinks-1, 2 and 3 respectively, it is sufficient that the following network alignment conditions are satisfied.

$$\text{Rank}[\hat{M}_{11}V_1 \quad \hat{M}_{21}V_2] = 2n+1$$
$$\Leftrightarrow \text{Rank}[V_1 \quad \hat{M}_{11}^{-1}\hat{M}_{21}V_2] = 2n+1 \quad (63)$$
$$\text{Rank}[\hat{M}_{22}V_2 \quad \hat{M}_{32}V_3] = 2n$$
$$\Leftrightarrow \text{Rank}[\hat{M}_{32}^{-1}\hat{M}_{22}V_2 \quad V_3] = 2n \quad (64)$$
$$\text{Rank}[\hat{M}_{13}V_1 \quad \hat{M}_{33}V_3] = 2n+1$$
$$\Leftrightarrow \text{Rank}[\hat{M}_{33}^{-1}\hat{M}_{13}V_1 \quad V_3] = 2n+1 \quad (65)$$

Suppose that (63)-(65) are satisfied. With $\underline{\gamma} = \{\gamma_{ij} \forall (i,j)\}$, let $f_1(\underline{\varepsilon}, \underline{\gamma}, \underline{\delta})$ denote the product of determinants of all the $2n \times 2n$ sub-matrices of $[\hat{M}_{32}^{-1}\hat{M}_{22}V_2 \quad V_3]$. Also, let

$$f_2(\underline{\varepsilon}, \underline{\theta}, \underline{\gamma}) = det([V_1 \quad \hat{M}_{11}^{-1}\hat{M}_{21}V_2])$$
$$f_3(\underline{\varepsilon}, \underline{\theta}, \underline{\delta}) = det([\hat{M}_{33}^{-1}\hat{M}_{13}V_1 \quad V_3])$$
$$f_4(\underline{\varepsilon}) = \prod_{(i,j)\in\{1,2,3\}|(i,j)\neq\{(3,1),(1,2),(2,3)\}} det(M_{ij}).$$

Let $f(\underline{\varepsilon}, \underline{\theta}, \underline{\gamma}, \underline{\delta}) = f_1(\underline{\varepsilon}, \underline{\gamma}, \underline{\delta})f_2(\underline{\varepsilon}, \underline{\theta}, \underline{\gamma})f_3(\underline{\varepsilon}, \underline{\theta}, \underline{\delta})f_4(\underline{\varepsilon})$. Since, $f_1(\underline{\varepsilon}, \underline{\gamma}, \underline{\delta})$, $f_2(\underline{\varepsilon}, \underline{\theta}, \underline{\gamma})$ and $f_3(\underline{\varepsilon}, \underline{\theta}, \underline{\delta})$ are non-zero polynomials, $f(\underline{\varepsilon}, \underline{\theta}, \underline{\gamma}, \underline{\delta})$ is also a non-zero polynomial. Hence, by Lemma 1 in [3] and Lemma 4, for a sufficiently large field size, there exists an assignment of values to the variables $\underline{\varepsilon}$, $\underline{\theta}$, $\underline{\gamma}$ and $\underline{\delta}$ such that the network alignment conditions are satisfied. Hence, the theorem is proved. ∎

## APPENDIX M
## PROOF OF THEOREM 9

*Proof:* To exactly recover $X_1'^{n+1}$, $X_2'^{n}$ and $X_3'^{2n+1}$ at the sinks-1, 2 and 3 respectively, it is sufficient that the following network alignment conditions are satisfied.

$$\text{Rank}[M_{33}] = 2n+1 \quad (66)$$
$$\text{Rank}[\hat{M}_{11}V_1 \quad \hat{M}_{21}V_2] = 2n+1$$
$$\Leftrightarrow \text{Rank}[V_1 \quad \hat{M}_{11}^{-1}\hat{M}_{21}V_2] = 2n+1 \quad (67)$$
$$\text{Rank}[\hat{M}_{22}V_2 \quad \hat{M}_{12}V_1] = 2n+1$$
$$\Leftrightarrow \text{Rank}[\hat{M}_{12}^{-1}\hat{M}_{22}V_2 \quad V_1] = 2n+1. \quad (68)$$

Since $\hat{M}_{33}$ is invertible (from Lemma 3), (66) is satisfied. Suppose that (67) and (68) are satisfied. Let

$$f_1(\underline{\varepsilon}) = det([M_{33}])$$
$$f_2(\underline{\varepsilon}, \underline{\theta}, \underline{\gamma}) = det([V_1 \quad \hat{M}_{11}^{-1}\hat{M}_{21}V_2])$$
$$f_3(\underline{\varepsilon}, \underline{\theta}, \underline{\gamma}) = det([\hat{M}_{12}^{-1}\hat{M}_{22}V_2 \quad V_1])$$
$$f_4(\underline{\varepsilon}) = \prod_{(i,j)\in\{1,2,3\}|(i,j)\neq\{(3,1),(3,2),(1,3),(2,3)\}} det(M_{ij}).$$

With $\underline{\gamma} = \{\gamma_{ij} \forall (i,j)\}$, let $f(\underline{\varepsilon}, \underline{\theta}, \underline{\gamma}) = f_1(\underline{\varepsilon})f_2(\underline{\varepsilon}, \underline{\theta}, \underline{\gamma})f_3(\underline{\varepsilon}, \underline{\theta}, \underline{\gamma})f_4(\underline{\varepsilon})$. Since, $f_1(\underline{\varepsilon})$, $f_2(\underline{\varepsilon}, \underline{\theta}, \underline{\gamma})$ and $f_3(\underline{\varepsilon}, \underline{\theta}, \underline{\gamma})$ are non-zero polynomials, $f(\underline{\varepsilon}, \underline{\theta}, \underline{\gamma})$ is also a non-zero polynomial. Hence, by Lemma 1 in [3] and Lemma 4, for a sufficiently large field size, there exists an assignment of values to the variables $\underline{\varepsilon}$, $\underline{\theta}$ and $\underline{\gamma}$ such that the network alignment conditions are satisfied. Hence, the theorem is proved. ∎